\documentclass{article}

\usepackage{arxiv}

\usepackage[utf8]{inputenc}
\usepackage[T1]{fontenc}

\usepackage{textcomp}
\usepackage{amsmath}
\usepackage{newtxtext}
\usepackage{newtxmath} 
\usepackage{bm}
\usepackage{graphicx}
\usepackage{booktabs}
\usepackage{array}
\usepackage{tabularx}
\usepackage{longtable}
\usepackage{enumitem}
\usepackage[font=small,labelfont=bf,labelsep=period]{caption}
\usepackage{xcolor}
\usepackage{microtype}
\usepackage[section]{placeins}
\usepackage{float}
\usepackage{cite}
\usepackage[hyphens]{url}
\usepackage{hyperref}
\makeatletter\g@addto@macro\UrlBreaks{\do\_}\makeatother
\hypersetup{colorlinks=true,linkcolor=blue!55!black,citecolor=blue!55!black,urlcolor=blue!55!black,
  pdftitle={Agentic Deep Elastic Strain Engineering of Semiconductor Electronic Properties},
  pdfauthor={Zhe Shi}}

\newcommand{\Eg}{\ensuremath{E_\mathrm{g}}}                 
\newcommand{\EgL}[1]{\ensuremath{E_\mathrm{g}(\mathrm{#1})}} 
\newcommand{\hunit}{meV\,\AA$^{-3}$}                         
\newcommand{\hval}[1]{\ensuremath{h = #1}\,\hunit}           
\newcommand{\cmi}{cm$^{-1}$}

\newcommand{\tool}[1]{\texttt{\detokenize{#1}}}             
\newcommand{\runid}[1]{\mbox{\detokenize{#1}}}             
\newcommand{\dir}[1]{\ensuremath{\langle #1\rangle}}       
\newcommand{\kgrid}[1]{\ensuremath{#1\times#1\times#1}}    
\newcommand{\strain}{\ensuremath{\bm{\varepsilon}}}         
\newcommand{\twopia}{\ensuremath{(2\pi/a_0)}}
\newcommand{\wmin}{\ensuremath{\omega_\mathrm{min}}}
\newcommand{\nq}{\ensuremath{n_q}}
\newcommand{\Dhh}{\ensuremath{D_{4h}}}
\newcommand{\Ddd}{\ensuremath{D_{3d}}}
\newcommand{\Oh}{\ensuremath{O_{h}}}

\renewcommand{\headeright}{Preprint}
\renewcommand{\undertitle}{Preprint}
\renewcommand{\shorttitle}{Agentic deep elastic strain engineering}

\title{Agentic Deep Elastic Strain Engineering of Semiconductor Electronic Properties}

\author{
  Zhe Shi\\
  Department of Materials Science and Engineering\\
  Massachusetts Institute of Technology\\
  Cambridge, MA 02139, USA\\
  \texttt{shizhe1@mit.edu}
}

\date{}

\begin{document}
\maketitle

\begin{abstract}
Deep elastic strain engineering (ESE) seeks the strain tensor that reaches a target electronic property at the least elastic energy. Machine-learning surrogates made this six-dimensional search tractable, but each campaign still needs an expert to pose the constrained problem and check the answer. We delegate that step to large-language-model (LLM) agents. A designer agent converts a one-sentence target into a constrained inverse problem and solves it through first-principles active learning on a valley-resolved, symmetry-equivariant surrogate. An independent verifier agent re-checks each candidate with valley-sampled hybrid-functional (HSE06) band edges and elastic stability. On pre-registered held-out tasks for Si and diamond, 8 of 10 agent runs and 8 of 9 runs of a scripted, LLM-free version of the same toolkit returned verified designs, whereas constrained Bayesian optimization, lacking a higher-fidelity check, never met the diamond 2\,eV target at full HSE06. The verified states include a multiaxial diamond state that reaches 2\,eV at about half the elastic energy of \dir{110} uniaxial tension, and a phonon-stable direct-gap Si state that needs 35--40\% less energy than the published triaxial route evaluated at the same level of theory. Because the LLM-free version matched the agent's success count, reliability in these runs came from the toolkit and the verification loop; the LLM's contribution was to turn a one-sentence goal into a correct constrained problem. Pairing a language model that formalizes the problem with deterministic search tools and an independent higher-fidelity verifier offers a template for verifiable autonomous inverse design beyond strain engineering.
\end{abstract}


\section{Introduction}\label{sec:intro}

Elastic strain is a continuous, reversible and six-dimensional control over the properties of a crystal, from its band structure and carrier transport to its thermal, magnetic and chemical behavior. In nanoscale and confined Si and in nanoscale diamond it can be pushed to strains of several percent before relaxation~\cite{zhang2016,banerjee2018,dang2021,nie2019,healy2014,grumstrup2014}: Si nanowires can be stretched repeatedly beyond 10\% elastic tensile strain, and individual wires reach about 16\%~\cite{zhang2016}; bent single-crystal diamond nanoneedles reach local tensile strains near 9\%~\cite{banerjee2018}; and microfabricated diamond bridges hold about 7.5\% uniform tension, rising to 9.7\% after geometry optimization~\cite{dang2021}. Strain at these scales also steers exciton transport in semiconductor microwires~\cite{fu2014} and couples mechanical motion to single spins in diamond~\cite{ovartchaiyapong2014}, and diamond anvils sustain multimegabar pressures~\cite{li2018dac,dubrovinsky2012}. These strains are an order of magnitude beyond the roughly 1\% of commercial strained-Si technology, so strain becomes a six-component design variable rather than a perturbation~\cite{li2014ese,zhu2010}. This regime is deep elastic strain engineering (ESE). The questions it poses are inverse ones: which strain tensor reaches a target property at the least elastic energy, and is that state physically admissible?

Machine-learning surrogates trained on first-principles data made such questions answerable over the full six-dimensional (6D) strain tensor. Neural networks trained on Perdew--Burke--Ernzerhof (PBE) data, fused with a smaller set computed in the GW approximation, predicted the electronic band gap to within about 8\,meV and exposed the indirect-to-direct and semiconductor-to-metal transitions of Si~\cite{shi2019}. That study also introduced the bandgap density of states and its lower envelope, which gives the least elastic energy density $h$ needed to reach a target gap. A physics-informed convolutional neural network later learned whole band structures, effective masses and Pareto fronts for diamond from about 35,000 PBE and 6,000 GW calculations, with active learning~\cite{tsymbalov2021}. Related machine-learning surrogates mapped the 6D phonon-stability boundary of diamond and the strain dependence of its lattice thermal conductivity~\cite{shi2024} and identified strain routes to metallize diamond~\cite{shi2020}.

Three difficulties remained. First, the gap landscape is non-smooth. The gap is the lowest of several conduction valleys ($\Delta$, L, $\Gamma$) measured from the valence-band maximum, so its isosurfaces carry ridgelines and corners where the conduction-band minimum (CBM) switches valley~\cite{shi2019}. A smooth regressor fitted directly to \Eg{} must learn these kinks from data: it needs many training points and is least accurate exactly at the valley crossings, where the direct-gap and semiconductor-to-metal transitions occur. Second, each campaign was expert-orchestrated: a specialist translated the physical goal into objectives and constraints, chose sampling domains, fidelity mixing, model architectures and active-learning rounds, and interpreted valley crossings by hand, so every new question meant rebuilding the pipeline. Third, physical admissibility and conventions were handled in separate, expert-run steps. Elastic and phonon stability were screened before or after the property calculations, for example with a separately learned phonon-stability boundary~\cite{shi2024}, rather than imposed inside the search, and the level of theory and the strain-measure conventions (Green--Lagrange or linear strain, tensor or engineering shear) were set by hand. A slip in such a convention yields a plausible number rather than an error message.

Both the computing and the expertise were costly: each study first built a global map from thousands of first-principles calculations, from which an expert read off targets~\cite{shi2019,tsymbalov2021}, translated them into constraints and checked the result. A user with a target property but without that expertise, such as an experimentalist, could not easily pose a new question. An agent that turns a one-sentence goal into a constrained problem and runs a directed search for it, without first building a global map, would lower this barrier, provided that its formalization and its answers can be checked.

Large-language-model (LLM) agents now operate density-functional-theory (DFT) workflows~\cite{liu2025vaspilot,harikumar2026,yang2026quasar}. They also predict and generate materials from language~\cite{kang2024}, acquire and accumulate their own research skills~\cite{huang2025cascade} and run closed-loop design, either directly over DFT~\cite{rothfarb2025} or with learned surrogate models~\cite{lu2025}. Closer to the present problem, invertible neural networks have inverse-designed the MoS$_2$ band gap under strain and field without a language model~\cite{fung2021}, and in a two-agent Creator--Inspector scheme for physics-constrained constitutive models, adding the inspecting agent raised the share of generated models that pass all physical-constraint checks~\cite{tacke2026}. LLMs used directly as Bayesian optimizers, by contrast, can ignore experimental feedback and lose to classical methods~\cite{gupta2025}. Two gaps remain: agents have not yet been applied to inverse design over a continuous strain tensor, and their reliability for such design is rarely measured.

This work addresses both gaps for deep ESE. First, we build an agentic framework in which a designer agent formalizes a plain-language target as an inverse problem over the full strain tensor and drives a first-principles campaign to solve it, and an independent verifier agent re-checks the answer. As language models used as Bayesian optimizers have not yet matched classical methods~\cite{gupta2025}, the search itself is left to a deterministic optimizer. Each campaign is a directed search for one goal rather than a global map: the 14 agent campaigns used 50--134 budget units each (one unit is one PBE calculation; Section~\ref{sec:framework}) and 3.6--146\,min of wall-clock time (Section~\ref{sec:agentcfg}). These costs are not comparable with those of a global map, which answers many questions at once; the point is that no map or hand-built pipeline is needed before the first question. Second, we evaluate the framework on pre-registered tasks and tolerances (Section~\ref{sec:tasks}) and ask two questions: does the agent turn natural-language ESE goals into correct constrained problems, and, once the problem is posed, does the language model add reliability beyond that of its toolkit and verification loop? We score the first with an automated rubric (Section~\ref{sec:fault}) and address the second with end-to-end campaigns compared with constrained Bayesian optimization (cBO) given an expert, the agent's or a naive formalization, and with a scripted, LLM-free run of the same toolkit (Section~\ref{sec:campaigns}).

\begin{figure}[H]
  \centering
  \includegraphics[width=\textwidth]{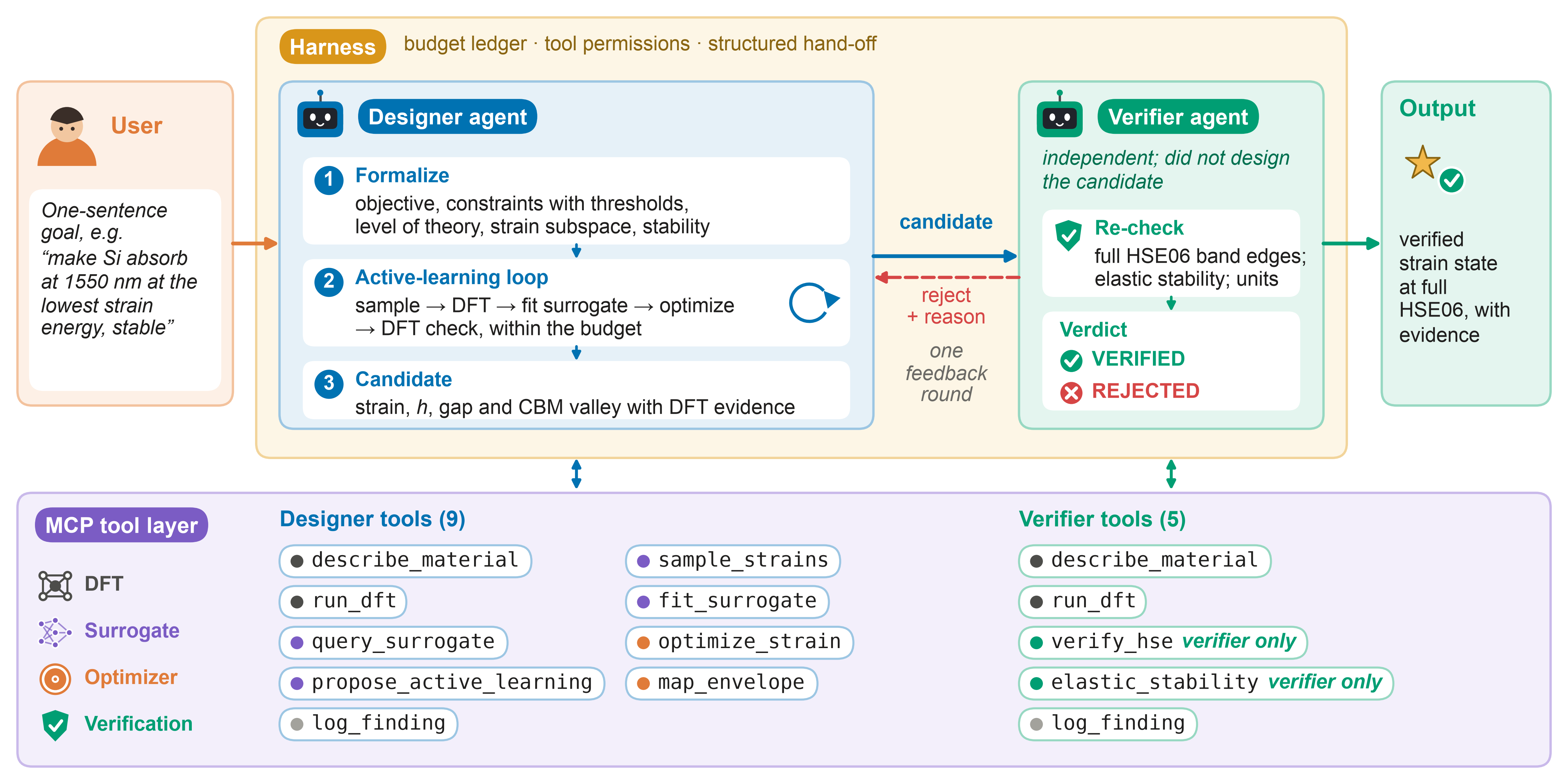}
  \caption{\textbf{Agentic deep-ESE framework.} The harness passes a one-sentence goal to the designer agent, which formalizes it, runs an active-learning campaign and hands one candidate to an independent verifier agent; a rejection returns to the designer with its reason (one feedback round). Both agents call a scoped Model Context Protocol (MCP) tool layer: \tool{verify_hse} and \tool{elastic_stability} are available to the verifier only (tool costs in Table~\ref{tab:tools}). The colors keep one meaning in all figures: blue, the agent and this work; green, verification and stability; red, rejection; purple, the MCP tools and the LLM judges; a gold star, a verified agent state.}
  \label{fig:framework}
\end{figure}

\section{Agentic framework}\label{sec:framework}

The framework runs two LLM agents, a designer and an independent verifier, inside a deterministic harness (Fig.~\ref{fig:framework}).

\begin{itemize}[leftmargin=*]
  \item \textbf{Harness.} It is deterministic code, not an agent: it hosts the two agents and owns the budget, isolation and control flow.
  \item \textbf{Designer (LLM agent).} Given a target property in natural language, it states the constrained inverse problem (objective, constraints, strain subspace, level of theory), runs the campaign that solves it and proposes a candidate strain.
  \item \textbf{Verifier (LLM agent).} It re-checks the designer's answer from first principles and alone holds the two tools that settle acceptance, full-HSE06 band edges (valley-sampled; Section~\ref{sec:absgap}) and elastic stability, and runs both on every candidate it accepts.
\end{itemize}

The designer and verifier run as separate headless (non-interactive, scripted) agent sessions, here in Claude Code (Section~\ref{sec:agentcfg}), although any agent runtime that runs headless and calls tools through the Model Context Protocol (MCP) would serve. The two sessions share no context: the verifier sees only the task, the designer's formalization and the candidate strain, never the designer's reasoning. A rejection returns the verifier's findings to the designer for one further round.

Both agents act only through an MCP server~\cite{mcp2024}, a standard interface that exposes typed functions to a language model. They have no file, shell or web access, which is confirmed from each session's initialization record. The tools are typed and unit-checked (Green--Lagrange strain, Voigt order, tensor shear, fractions) and are listed in Table~\ref{tab:tools}.

\begin{table}[H]
  \centering
  \caption{MCP tools available to the designer and verifier agents, with their cost in budget units (1 unit = one PBE calculation; 150 units per run, shared by both agents). Tools that cost 0 run no DFT.}
  \label{tab:tools}
  \small
  \begin{tabularx}{\textwidth}{@{}l X c c l@{}}
    \toprule
    \textbf{Tool} & \textbf{Function} & \textbf{Designer} & \textbf{Verifier} & \textbf{Cost (units)} \\
    \midrule
    \tool{describe_material} & reference data, conventions, costs and remaining budget & \checkmark & \checkmark & 0 \\
    \tool{sample_strains} & Latin-hypercube strain samples (no DFT) & \checkmark & & 0 \\
    \tool{run_dft} & PBE DFT with valley-resolved band edges and $h$ & \checkmark & \checkmark & 1 per PBE point \\
    \tool{fit_surrogate} & fits the valley-resolved surrogate to the run's DFT points & \checkmark & & 0 \\
    \tool{query_surrogate} & surrogate gap, its uncertainty, $h$ and CBM valley & \checkmark & & 0 \\
    \tool{propose_active_learning} & proposes new strains for DFT (Section~\ref{sec:surrogate}) & \checkmark & & 0 \\
    \tool{optimize_strain} & deterministic multi-start constrained optimizer & \checkmark & & 0 \\
    \tool{map_envelope} & surrogate lower and upper envelopes of the gap vs $h$ & \checkmark & & 0 \\
    \tool{verify_hse} & full-HSE06 band edges at one strain & & \checkmark & 15 \\
    \tool{elastic_stability} & elastic-stability criteria from DFT stresses & & \checkmark & 12 \\
    \tool{log_finding} & appends a finding to the run's notebook & \checkmark & \checkmark & 0 \\
    \bottomrule
  \end{tabularx}
\end{table}

The language model never does the numerical search: it chooses the objective, constraints, subspace, bounds, robustness margin and level of theory, and \tool{optimize_strain} solves the problem the language model posed.

Every tool call that runs a calculation is charged against a fixed budget of 150 units per run (one unit is the cost of one PBE calculation), shared by designer and verifier (costs in Table~\ref{tab:tools}). A content-addressed cache avoids recomputing identical states. A cache hit is charged like a miss and returns identical output, so the cache changes neither a run's charges nor its results. Each run has its own data store, ledger and notebook, and its full transcript is kept, including every tool call and result.

\section{Methods}\label{sec:methods}

\subsection{First-principles calculation settings}\label{sec:dft}
We use Quantum ESPRESSO 7.6~\cite{giannozzi2009,giannozzi2017} with the PBE functional~\cite{perdew1996} and PseudoDojo norm-conserving pseudopotentials~\cite{vansetten2018}. The cutoffs are 50\,Ry for Si and 100\,Ry for diamond. Calculations use two-atom primitive cells and an \kgrid{8} Monkhorst--Pack $k$-mesh (\kgrid{10} for the benchmark sets of Section~\ref{sec:surr-results}; convergence tests in Supplementary Section~\ref{sec:S9}). Internal coordinates are relaxed in the fixed strained cell. Occupations use Marzari--Vanderbilt smearing of 0.005\,Ry (0.002\,Ry in HSE06 calculations), so metallic states such as those of tasks T1 and H3 (Table~\ref{tab:tasks}) are well defined. Strain is the Green--Lagrange strain tensor \strain{}, applied through the rotation-free deformation gradient $\bm{F} = (\bm{I} + 2\strain)^{1/2}$, and the elastic energy density is $h = [E(\strain) - E_0]/V_0$, where $E$ is the total energy of the cell and $E_0$ and $V_0$ are the energy and volume of the unstrained cell. As validation, PBE gives a lattice parameter $a_0 = 5.4689$\,\AA{} and a bulk modulus $B = 88.3$\,GPa for Si, and 3.5724\,\AA{} and 432.3\,GPa for diamond.

\subsection{Valley-resolved band edges}\label{sec:valley}
Conduction-band valleys are tracked in the unstrained cubic frame. Reduced $k$-coordinates are invariant under homogeneous strain, so the same labels apply at every strain. $\Gamma$ and the four L points are time-reversal-invariant momenta and are read directly; each $\Delta$ valley is located on its $\Gamma$--X line and then refined on a local stencil. The gap is the lowest conduction-valley energy minus the valence-band maximum, and a uniform-grid check flags any minimum outside the tracked valleys. At zero strain the Si CBM lies at 0.843\,\twopia{} along $\Gamma$--X, with $\EgL{PBE} = 0.607$\,eV, and the diamond CBM at 0.723\,\twopia{}, with $\EgL{PBE} = 4.173$\,eV.

\subsection{Absolute gaps}\label{sec:absgap}
This work tests whether an agent can pose and run a deep-ESE campaign correctly, not how accurately gaps can be computed; earlier deep-ESE studies already reached GW-level gap accuracy with machine-learning surrogates~\cite{shi2019,tsymbalov2021}. We therefore use an inexpensive, transparent protocol whose known error is itself part of the test, because the verifier has to catch it. PBE underestimates gaps, so absolute targets use a rigid per-valley shift, the ``HSE scissor'', taken from zero-strain HSE06 calculations~\cite{heyd2003,krukau2006} ($\nq = 4$, i.e.\ a \kgrid{4} $q$-mesh for Fock exchange; \kgrid{8} $k$-mesh) and used during search. For Si the shifts are $\Gamma$~+0.764\,eV, L~+0.763\,eV and $\Delta$~+0.582\,eV, giving $\EgL{HSE06} = 1.189$\,eV; for diamond they are $\Gamma$~+1.391\,eV, L~+1.562\,eV and $\Delta$~+1.099\,eV, giving 5.272\,eV. A gap obtained this way is the HSE-scissor gap. Critical states are re-evaluated with HSE06 at their own strain, on an \kgrid{8} $k$-mesh without symmetry, with $\nq = 4$, a Fock-exchange cutoff twice the wavefunction cutoff and a self-consistent-field (SCF) threshold of $10^{-9}$\,Ry. The comparison arms used $10^{-7}$\,Ry and an equal Fock cutoff, a difference that changes gaps by less than 1\,meV. The HSE06\,$-$\,PBE difference is evaluated at grid points of each valley ($\Gamma$, L and 0.75\,\twopia{} for $\Delta$) and added to the PBE valley minima. ``Full HSE06'' in this paper means this valley-sampled evaluation, neither a dense-mesh HSE06 band structure nor a search for the band extrema at the HSE06 level; its difference from the HSE-scissor gap is the scissor error. Taking the $\Delta$ correction at 2/3 or 5/6 of the way along $\Gamma$--X on a \kgrid{12} grid instead of at 0.75 changes the gaps of the H1 agent state and the bracketing \dir{110} states by at most 0.03\,eV, all in the same direction (1.972--2.004\,eV for the H1 agent state). In the strained diamond states of H1 the PBE minimum of the lowest $\Delta$ valley lies at X; taking the correction there lowers the gaps of two of the three accepted H1 states to 1.962\,eV (agent) and 1.964\,eV (scripted), 6--8\,meV below the window, and changes the verdict of no other state checked: the other agent and scripted H1 states (including the H1 reference, 1.994\,eV), the cBO-agent H1 state and every Si T1 and H3 state (Supplementary Section~\ref{sec:S10}). Scoring keeps the evaluation at 0.75\,\twopia{} used throughout; scored at X, the held-out counts would be 7 of 10 and 7 of 9. The $\pm 0.03$\,eV window of H1 is a tolerance of this protocol, not an estimate of its accuracy.

\subsection{Surrogate}\label{sec:surrogate}
One neural-network ensemble per valley family is evaluated in that valley's own symmetry frame. Each ensemble has five fully connected networks (three hidden layers of 64 SiLU units), trained with Adam for 600 epochs (cosine-annealed learning rate from $5\times10^{-3}$) on strains scaled by 20; a separate ensemble predicts $h$. Invariance under the valley's stabilizer group is exact through group averaging (\Dhh{} for $\Delta$, \Ddd{} for L, \Oh{} for $\Gamma$ and for $h$). The inputs include the Bir--Pikus invariants~\cite{bir1974}, which let the $\Gamma$ networks follow the conical splitting of the degenerate $\Gamma$ states, although the split branches are still not captured accurately near their crossing (Section~\ref{sec:surr-results}). The gap is the minimum over predicted valleys, with uncertainty from the ensemble spread. The baselines are an otherwise identical ensemble that regresses \Eg{} directly and a simplified band-on-grid network that predicts the lowest conduction band at 145 fixed $k$-points (a \kgrid{6} grid plus the $\Gamma$--X lines) and the valence-band maximum. The latter borrows the band-on-grid idea of Tsymbalov et~al.~\cite{tsymbalov2021} but not their convolutional architecture, which was built for tens of thousands of calculations, far more than a campaign budget. All three models share a training schedule, so they differ mainly in what they predict. They are compared on three DFT benchmark sets, each of 320 Latin-hypercube Green--Lagrange strains with normal components within $\pm$3\% and shear components within $\pm$2\% (Si, small strain; 307 converged states), within $\pm$10\% and $\pm$6\% (Si, wide range; 319) and within $\pm$10\% and $\pm$8\% (diamond; 316). Errors are taken on the signed gap, which is negative for metallic states; 296 of the 319 wide-range Si states are metallic at PBE.

\paragraph{Active learning.} The designer can request new DFT points through \tool{propose_active_learning}. The tool draws 4,000 candidate strains within the bounds and scores each by the ensemble spread of the predicted gap, divided by $[1 + h/(5\,\text{\hunit})]$ to favor low-energy states. When a gap constraint is given, the score is multiplied by $\exp(-|\Eg - E_{\mathrm{g,target}}|/(0.15\,\mathrm{eV}))$ to concentrate sampling near the constraint boundary. It returns the top-scoring strains that are at least 0.01 apart. Whether to use the proposals, how many to compute and when to refit the surrogate are the agent's decisions.

\subsection{Stability}\label{sec:stability}
Elastic stability at a strained state is tested from DFT stresses at $\pm$0.5\% perturbations of each Green--Lagrange strain component. At a prestressed state the work-conjugate pair is the second Piola--Kirchhoff stress $\bm{S} = J\,\bm{F}^{-1}\bm{\sigma}\bm{F}^{-\mathsf{T}}$, with $J = \det\bm{F}$ and $\bm{\sigma}$ the Cauchy stress, and the Green--Lagrange strain \strain{}, and we use positive definiteness of the material tangent $\partial\bm{S}/\partial\strain$ as a screening criterion for elastic stability under homogeneous deformation. At zero stress this criterion reduces to the necessary and sufficient criteria of Mouhat and Coudert~\cite{mouhat2014}. At finite prestress it does not by itself establish stability under every loading condition, because the second variation then depends on the permitted perturbations and on the loading~\cite{wang1995}. The campaign tool used the Cauchy stress in place of $\bm{S}$; re-evaluating all 35 states afterwards with a full stress stencil (23 Si, 12 diamond) from the same DFT stresses using $\bm{S}$ changed no acceptance verdict (Supplementary Table~\ref{tab:S2}); the minimum eigenvalues are 11.3--74.9\,GPa for Si and above 108\,GPa for diamond. Values in the text and Fig.~\ref{fig:physics}b come from this reanalysis; transcript quotes (Fig.~\ref{fig:h2}a) come from the original check. Phonon stability is computed with density-functional perturbation theory (DFPT)~\cite{baroni2001} (\texttt{ph.x}, \kgrid{4} $q$-grid, Fourier-interpolated to \kgrid{16}) only at the reported states where it is claimed.

\subsection{Tasks, arms and scoring}\label{sec:tasks}
Tasks, tolerances, the scoring rubric and the cBO comparison arms were pre-registered and committed before any agent run; the scripted arm was added after the agent runs, in a dated amendment recorded before any scripted run. Each task is given to the agent as the verbatim sentence in Table~\ref{tab:tasks}. Every later change is listed in Supplementary Table~\ref{tab:S2}, with its amendment date where one was recorded.

\begin{table}[!tbp]
  \centering
  \caption{Tasks given to the agent and pre-registered success criteria. T1 and T2 are calibration tasks with published answers; H1--H3 are held-out tasks. \Eg{} is the gap at full HSE06 (valley-sampled; Section~\ref{sec:absgap}); $\varepsilon_{33}$ is the strain normal to the (001) film. Prompts are reproduced verbatim.}
  \label{tab:tasks}
  \small
  \begin{tabularx}{\textwidth}{@{}l l >{\raggedright\arraybackslash}X >{\raggedright\arraybackslash}p{3.6cm}@{}}
    \toprule
    \textbf{ID} & \textbf{Class} & \textbf{Prompt given to the agent (verbatim)} & \textbf{Success criterion} \\
    \midrule
    T1 & calibration & ``Find the strain state that turns silicon into a metal (zero band gap) at the lowest possible elastic strain energy.'' & $\Eg \le 0$ at HSE06 \\
    \addlinespace
    T2 & calibration & ``Find the strain state that makes silicon a direct-band-gap semiconductor at the lowest possible elastic strain energy, and assess whether that state is stable.'' & $\Gamma$ is the conduction-band minimum by $\ge 0.01$\,eV \\
    \addlinespace
    H1 & held-out & ``Find the strain state that lowers the band gap of diamond to 2.0 eV at the lowest possible elastic strain energy, and compare it with simple uniaxial loading along \textless100\textgreater, \textless110\textgreater{} and \textless111\textgreater.'' & $|\Eg - 2.0\,\mathrm{eV}| \le 0.03\,\mathrm{eV}$ at HSE06 \\
    \addlinespace
    H2 & held-out & ``Make silicon absorb light at 1550 nm at the lowest possible elastic strain energy, while keeping the crystal stable.'' & $\Eg \le 0.80$\,eV at HSE06, elastically stable \\
    \addlinespace
    H3 & held-out & ``A silicon film is constrained so that its normal strain perpendicular to the (001) surface is zero. Find the lowest-elastic-energy strain state that closes its band gap.'' & $\Eg \le 0$ at HSE06 with $\varepsilon_{33} = 0$ \\
    \bottomrule
  \end{tabularx}
\end{table}

Calibration tasks have published answers; held-out tasks do not. Run labels give the task, the model and the replicate: s = Claude Sonnet, o = Claude Opus (e.g.\ \runid{H2_s3}).

Each held-out task is attempted by three methods, which give six comparison arms:
\begin{itemize}[leftmargin=*]
  \item \textbf{the agent} (designer plus verifier), with 150 units;
  \item \textbf{cBO}, constrained Bayesian optimization with expected improvement~\cite{gardner2014}, implemented with BoTorch 0.17~\cite{balandat2020}, run directly on DFT with 100 units. It uses one Gaussian process per objective and constraint (Mat\'ern-5/2, automatic relevance determination), 20 Latin-hypercube initial points, batches of 10 chosen by expected improvement weighted by the probability of feasibility, maximized over 4,000 random candidates, and strain bounds of $\pm$0.10. cBO is given three formalizations, which define the arms cBO-expert, cBO-agent and cBO-naive:
  \begin{itemize}
    \item the expert's, written by the author before any agent run;
    \item the agent's own, taken from its log;
    \item a naive one, fixed in advance: raw-PBE thresholds, no stability constraint, full 6D strain;
  \end{itemize}
  \item \textbf{the scripted arm}, an LLM-free run of the agent's tools with the expert formalization (at the HSE-scissor level) and a fixed schedule: DFT at 20 Latin-hypercube strains, then three rounds of \tool{fit_surrogate}, \tool{optimize_strain} (bound 0.10) and DFT at the top three candidates. The lowest-$h$ computed state that meets the constraint is evaluated with \tool{verify_hse} and \tool{elastic_stability} and accepted in code, not by a verifier agent, if it meets the scoring rule below. A failed state triggers one re-target, mirroring the verifier's feedback round: the gap window is shifted by the search-level gap minus the full-HSE06 gap at that state and one more round is run, or, if the gap criterion was met, the next-best state is taken. Three seeds per task; after the campaigns it was also run once per task with the naive formalization (dated amendment; Supplementary Section~\ref{sec:S10}).
\end{itemize}

The request for the H2 formalization included an example that closely matched the answer given, so the H2 expert formalization is not blind. Phonon stability, required by the expert formalizations, is not enforced inside any search; it is checked afterwards at reported states. cBO's final states are scored with HSE06 outside its 100 units.

All arms are scored alike: a state succeeds if it meets the Table~\ref{tab:tasks} criterion at HSE06 and is elastically stable. Elastic stability is thus required on every task, not only on H2, whose prompt asks for it; every state that met its gap criterion was elastically stable, so no outcome depends on this choice. For each task, the reference $h_\mathrm{ref}$ is the lowest $h$ among successful states from any arm, agent included. No separate search for the true optimum was run, so the reference is an upper bound on the optimum. Regret is the excess elastic energy density $h - h_\mathrm{ref}$ over the task reference, in \hunit.

\subsection{Agent configuration}\label{sec:agentcfg}
Designer and verifier run in Claude Code 2.1.305 as \texttt{claude-sonnet-5-5} (s runs) or \texttt{claude-opus-5-5} (o run). The closed-book probe (Section~\ref{sec:probe}), the formalization study and the fault-injection replays (Section~\ref{sec:fault}) ran Claude Sonnet single-turn with tools off, the replays with the verifier's system prompt; the automated judges, Claude Sonnet and Claude Haiku, ran the same way. System prompts and the judges' rubric and answer key are given in Supplementary Sections~\ref{sec:S1} and~\ref{sec:S3}; full transcripts are available from the author on request. Wall-clock time per run (3.6--146\,min) and token use are listed in Supplementary Section~\ref{sec:S6}.

\section{Results}\label{sec:results}

\subsection{A valley-resolved surrogate needs far fewer DFT points}\label{sec:surr-results}

\begin{figure}[!tbp]
  \centering
  \includegraphics[width=\textwidth]{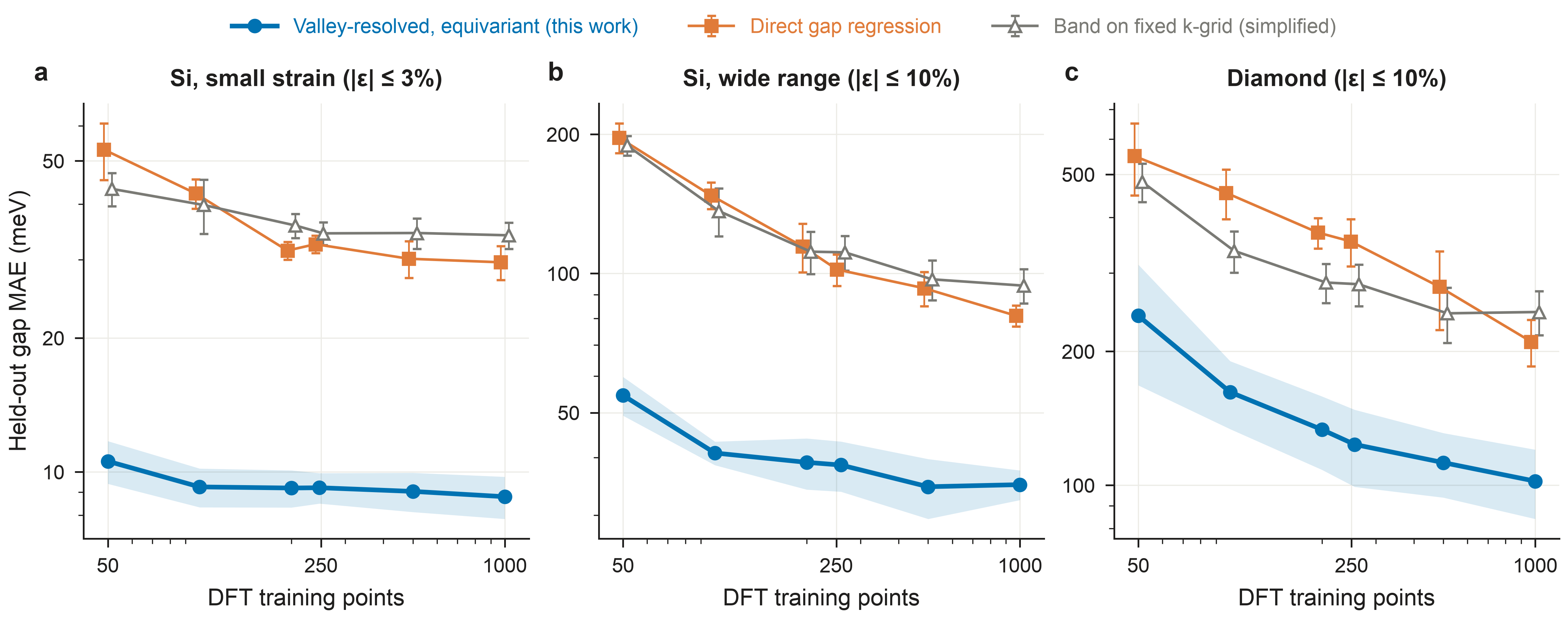}
  \caption{\textbf{Valley-resolved surrogates need far fewer DFT points.} Held-out gap mean absolute error (MAE, meV) vs number of DFT training points for three surrogates on three DFT benchmark sets: (a)~Si at small strain ($|\varepsilon_{ii}| \le 3\%$, $i = 1,2,3$), (b)~Si over a wide range ($|\varepsilon_{ii}| \le 10\%$, $i = 1,2,3$) and (c)~diamond ($|\varepsilon_{ii}| \le 10\%$, $i = 1,2,3$). Training sets larger than the pre-campaign training sets (247, 259 and 256 points) add states from the same sampling boxes, computed after the campaigns (Supplementary Tables~\ref{tab:S4} and~\ref{tab:S7}); both axes are logarithmic. Lines are means over five random train--test splits; the shaded band (this work) and the error bars (baselines, offset by $\pm$3\% in training size for legibility) are $\pm$1 standard deviation. Identical training and test data for all models; the band-on-grid model is a simplified fixed-grid baseline, not the convolutional model of ref.~\cite{tsymbalov2021}. The diamond MAE exceeds the 0.03\,eV H1 tolerance at all training sizes. The larger exploratory network of Table~\ref{tab:mae} is not shown.}
  \label{fig:surrogate}
\end{figure}

Is the designer's main instrument accurate enough to steer a search on a small budget? On every DFT benchmark set, predicting the smooth valley energies and taking their minimum beats regressing the gap directly (Fig.~\ref{fig:surrogate}, Table~\ref{tab:mae}). With the same training data, the valley-resolved error is 2.3--3.7$\times$ lower than that of both baselines, direct gap regression and the band-on-grid model (Section~\ref{sec:surrogate}). The gain matters most under a campaign budget: for both Si sets the valley-resolved model trained on only 50 points is more accurate than either baseline trained on 1,000 points, twenty times as many (Fig.~\ref{fig:surrogate}). It also identifies the CBM valley correctly in 97--99\% of held-out Si states and in 90--95\% of diamond states.

\begin{table}[!tbp]
  \centering
  \caption{Held-out gap mean absolute error (MAE, meV) of the surrogates on three DFT benchmark sets, mean over five random train--test splits with 60 held-out states each. Strain ranges are the sampling bounds on the normal strain components; the shear bounds are 2\%, 6\% and 8\% (Section~\ref{sec:surrogate}).}
  \label{tab:mae}
  \small
  \begin{tabularx}{\textwidth}{@{}l *{5}{>{\centering\arraybackslash}X}@{}}
    \toprule
    \textbf{Benchmark set} & \textbf{Valley-resolved (used in campaigns)} & \textbf{Direct regression} & \textbf{Band on grid (simplified)} & \textbf{Valley-resolved, 50 points} & \textbf{Valley-resolved, larger network (exploratory)} \\
    \midrule
    Si, small strain ($|\varepsilon_{ii}| \le 3\%$) & 9.2 & 32.5 & 34.4 & 10.6 & 3.8 \\
    Si, wide range ($|\varepsilon_{ii}| \le 10\%$) & 38.5 & 102.0 & 111.1 & 54.5 & 24.6 \\
    Diamond ($|\varepsilon_{ii}| \le 10\%$) & 123.6 & 353.5 & 283.2 & 240.6 & 77.6 \\
    \bottomrule
  \end{tabularx}
  \par\smallskip
  \begin{minipage}{\textwidth}\footnotesize
  Split-to-split standard deviations are in Supplementary Table~\ref{tab:S3}. The first three model columns use the full pre-campaign training sets (247, 259 and 256 points). The last column is the larger network of an exploratory study after the campaigns (Supplementary Section~\ref{sec:S2}), not used in them.
  \end{minipage}
\end{table}

The surrogate can therefore steer a search but not certify a state. For Si its error flattens beyond about 100 points and changes little up to 1,000 (8.8 and 34.9\,meV). For diamond it is about 124\,meV at 256 points; states added from the same sampling box after the campaigns lower it only to about 102\,meV at 1,000 points and not further up to 6,456 points (Supplementary Table~\ref{tab:S4}), still far above the 0.03\,eV H1 tolerance. Every reported state is therefore checked by DFT and full HSE06 (Sections~\ref{sec:framework} and~\ref{sec:absgap}). Most of the residual error comes from states whose valence-band maximum leaves $\Gamma$ at extreme strain and from diamond states whose conduction-band minimum is the degenerate $\Gamma$ valley, which the valley representation cannot express; the larger network of Table~\ref{tab:mae} lowers the error 1.6--2.4$\times$ but was not used in the campaigns (Supplementary Sections~\ref{sec:S2} and~\ref{sec:S8}).

\subsection{Closed-book contamination probe}\label{sec:probe}

Could the model simply recall the published calibration answers? We asked Claude Sonnet, the model used for the calibration runs, the T1 and T2 questions with no tools, ten times each. It never reproduced the published values (1.35 or 1.65\,\hunit{} for Si metallization; 15.4\,\hunit{} or 9.3\% for direct-gap Si), nor the strain tensor as published in 2019, before its 2020 correction~\cite{shi2019}, a canary that only verbatim recall could reproduce. Its guesses for $h$ ranged from about 1 to 100\,\hunit, and it said it did not remember the result; the agent's own T2 state also differs from the published one (Section~\ref{sec:physics}). The probe cannot exclude implicit priors, such as knowing that tension lowers the $\Gamma$ valley, so T1 and T2 serve as checks that the method recovers known behavior, not as discoveries.

\subsection{Held-out inverse-design campaigns}\label{sec:campaigns}

\begin{figure}[!tbp]
  \centering
  \includegraphics[width=\textwidth]{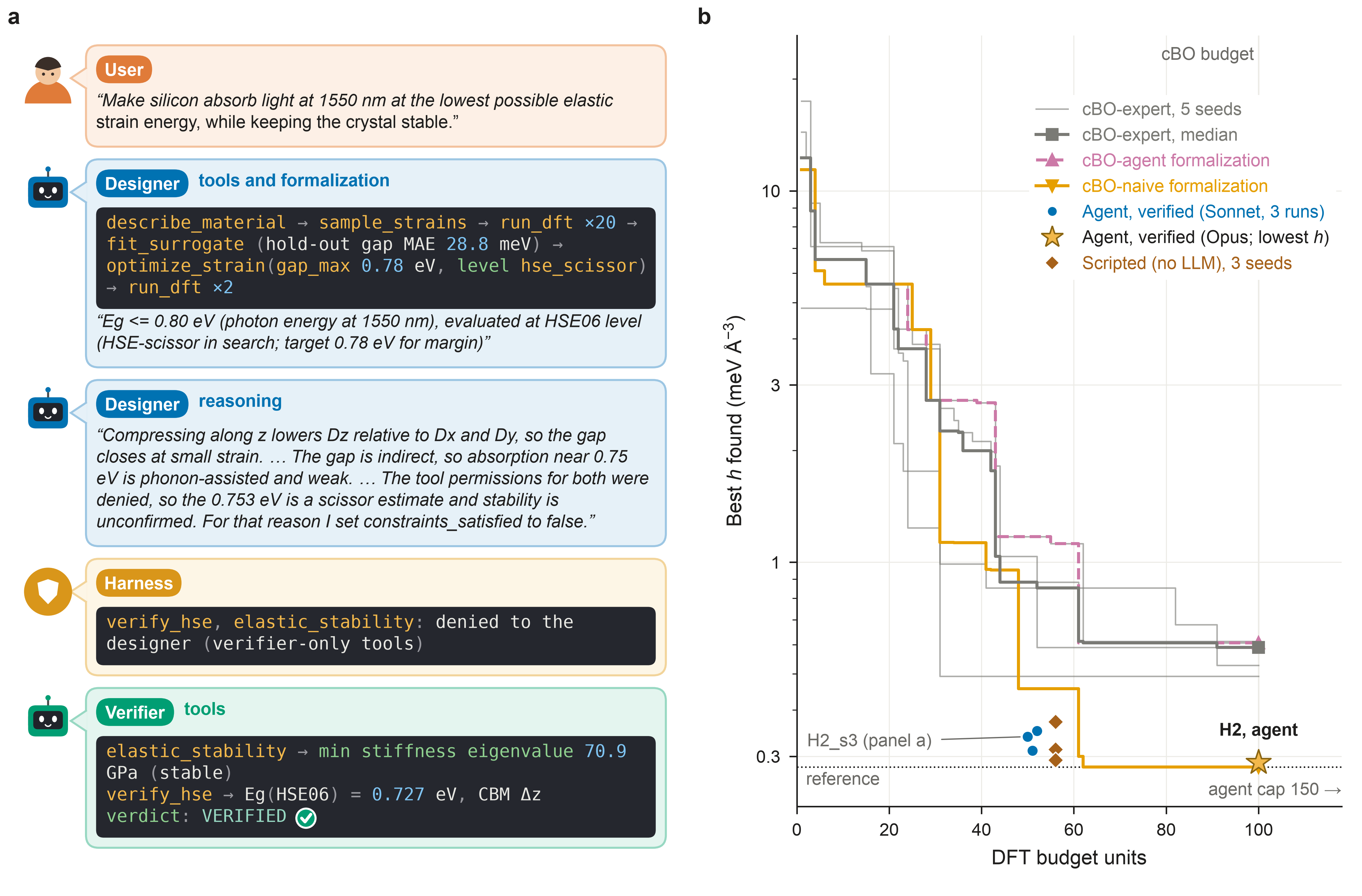}
  \caption{\textbf{An example campaign and budget efficiency on H2.} (a)~Verbatim excerpt of \runid{H2_s3}, the median run under the amended selection rule (Supplementary Table~\ref{tab:S2}). Omissions are marked \ldots; tool calls are summarized by name; Dz denotes the $\Delta$ valley along $z$. The stiffness eigenvalue in the verifier output is from the original Cauchy-stress form of the stability check (Section~\ref{sec:stability}). The designer set \tool{constraints_satisfied} to false because it could not run the verifier's tools; the verifier then ran them and accepted the state against the pre-registered gap and stability criteria. (b)~Best $h$ found (\hunit) vs DFT budget units (log $y$ axis) for cBO with the expert (5 seeds, thin lines; bold line, their median), the agent's and the naive formalizations; cBO curves use the HSE-scissor gap and stop at the 100-unit cBO budget (light vertical line), whereas the agents' cap was 150 units. Circles are the verifier-accepted states of the three Sonnet runs, the star that of the Opus run (the lowest agent $h$ on H2), and diamonds those of the scripted-arm runs, all at their total spend; the dotted line is the task reference (Section~\ref{sec:tasks}).}
  \label{fig:h2}
\end{figure}

\begin{figure}[!tbp]
  \centering
  \includegraphics[width=\textwidth]{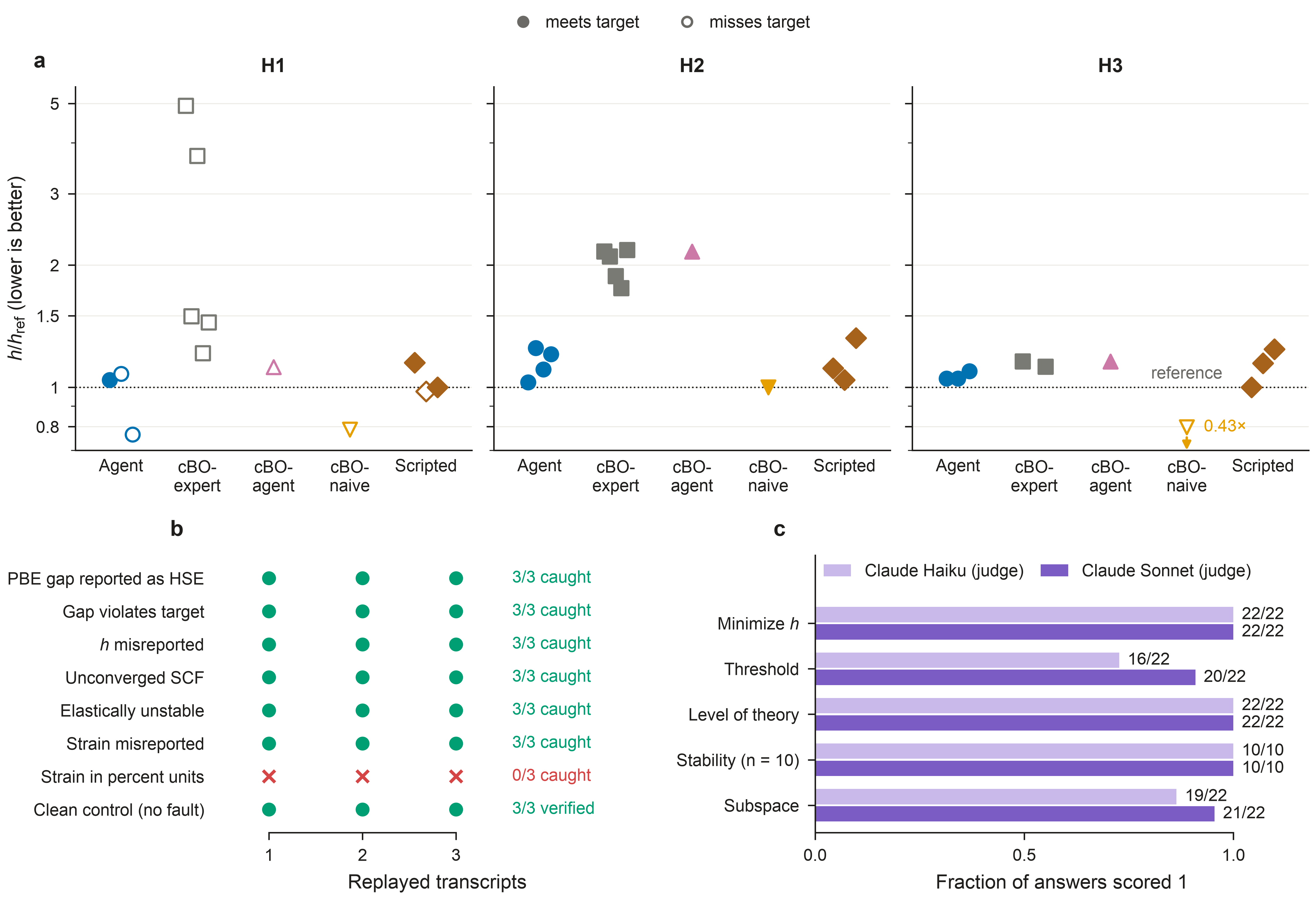}
  \caption{\textbf{Outcomes of all arms and single-turn evaluations.} (a)~Outcomes on the held-out tasks for the agent, for cBO with the expert, agent's and naive formalizations, and for the scripted arm, as $h$ relative to the task reference ($h_\mathrm{ref} = 29.103$, 0.281 and 2.331\,\hunit{} for H1, H2 and H3), on a log axis shared by the three panels. Each arm has a fixed color and marker. H1 cBO-expert and cBO-naive columns show the corrected two-sided-window rerun; the H3 cBO-naive state ($h/h_\mathrm{ref} = 0.43$) lies below the axis range and is drawn at the bottom edge with an arrow; it is not metallic at full HSE06 and lies outside the $\varepsilon_{33} = 0$ subspace. Filled markers meet the target at full HSE06 and are elastically stable (for the agent, verifier-accepted; for the scripted arm, accepted by the same checks in code; for cBO, checked after the campaigns with the same criterion); open markers fail, and an open marker below 1 is not a better solution. The scripted arm with the naive formalization is reported in Table~\ref{tab:outcomes}. (b)~Verifier outcomes on replayed transcripts with one injected fault each: filled circles are correct verdicts (fault rejected, or clean control verified); $\times$ are misses. The strain-in-percent-units fault was missed by the verifier but is blocked upstream by the tool schema ($|\varepsilon_{ij}|$ bounds). (c)~Fraction of the 22 formalization answers scored 1 on each rubric criterion by two automated judges (Claude Sonnet and Haiku), with counts at the bar ends; stability is scored only for the 10 goals that ask for it. Inter-judge $\kappa = 0.48$ (raw agreement 94\%).}
  \label{fig:outcomes}
\end{figure}

Does the designer--verifier team deliver on tasks without published answers, and if so, who deserves the credit? For each held-out task the designer formalized the goal sentence, ran the campaign and handed one candidate to the verifier (Fig.~\ref{fig:h2}a). The outcomes are compared with the five other arms of Section~\ref{sec:tasks} (Table~\ref{tab:outcomes}, Fig.~\ref{fig:outcomes}a). The key control is the scripted arm, which has everything the agent has except the language model. Run counts are small, so we report counts and case studies, not success rates.

\begin{table}[!tbp]
  \centering
  \caption{Held-out outcomes by arm.}
  \label{tab:outcomes}
  \small
  \setlength{\tabcolsep}{4pt}
  \begin{tabularx}{\textwidth}{@{}l *{6}{>{\centering\arraybackslash}X}@{}}
    \toprule
    \textbf{Task} & \textbf{Agent} & \textbf{Scripted arm} & \textbf{cBO-expert} & \textbf{cBO-agent} & \textbf{cBO-naive} & \textbf{Scripted, naive} \\
    \midrule
    H1 & 1/3 (30.334) & 2/3 (\textbf{29.103}, 33.507) & 0/5 & 0/1 & 0/1 & 0/1 \\
    \addlinespace
    H2 & 4/4 (0.289--0.351) & 3/3 (0.293--0.372) & 5/5 (0.493--0.612) & 1/1 (0.607) & 1/1 (\textbf{0.281}) & 1/1 (0.36) \\
    \addlinespace
    H3 & 3/3 (2.451--2.557) & 3/3 (\textbf{2.331}--2.894) & 2/2 (2.618, 2.698) & 1/1 (2.698) & 0/1 (not metallic; $\varepsilon_{33} \neq 0$) & 0/1 (not metallic; $\varepsilon_{33} \neq 0$) \\
    \bottomrule
  \end{tabularx}
  \par\smallskip
  \begin{minipage}{\textwidth}\footnotesize
  Successful runs out of all runs, with the $h$ of the successful states (\hunit) in parentheses; bold marks the task reference $h_\mathrm{ref}$. Agent runs succeed when the verifier accepts them, scripted-arm runs when the same full-HSE06 and stability checks pass in code; cBO runs when they meet the target at full HSE06 and are elastically stable (checked after the campaigns with the verifier's criterion; minimum stiffness eigenvalues 62.3--70.9\,GPa on H2 and 25.4--25.8\,GPa on H3). H2 budget spent: agent 50--52 units (Sonnet) and 100 (Opus); scripted arm 56 units. H1 cBO-expert and cBO-naive entries are the two-sided-window rerun. On H2 and H3 the cBO-agent run reached the same state as cBO-expert seed 0. The scripted, naive column (one seed per task, run after the campaigns) gives the scripted arm the naive formalization with the same tools, budget, schedule and verification checks. Budgets were 150 units per run for the agent and the scripted arm and 100 units for cBO, whose final HSE06 scoring was not charged. Scored with the $\Delta$ correction at the valley minimum, H1 would be 0/3 for the agent and 1/3 for the scripted arm (Section~\ref{sec:absgap}).
  \end{minipage}
\end{table}

\textbf{The team delivers, and the credit goes to the toolkit and the verification loop.} Eight of ten agent runs were verified, against eight of nine runs of the scripted arm, which shares the agent's tools, budget, surrogate and verification calculations (full HSE06 and elastic stability, judged in code rather than by a verifier agent) but is handed the expert's formalization and follows a fixed schedule (Section~\ref{sec:tasks}). \runid{H3_s2} repeated \runid{H3_s1} exactly (Section~\ref{sec:limits}), so the ten agent runs contain nine distinct trajectories, seven of them verified. The scripted states have comparable $h$, at similar or lower spend. Once the problem is posed, the language model's search decisions therefore add no measurable reliability. Holding the toolkit and verification checks fixed but giving the scripted arm the naive formalization isolates the formalization: it then passed only H2, after its re-target, and failed H1, landing at 1.933\,eV after re-targeting from a raw-PBE window, and H3, whose state was not metallic at full HSE06 and lay outside the H3 subspace ($\varepsilon_{33} = -4.2\%$) (Table~\ref{tab:outcomes}; Supplementary Section~\ref{sec:S10}). The cBO arms, which search without a higher-fidelity check, all failed where the inexpensive gap estimate misled them (H1), and cBO-expert reached 1.8--2.2 times the reference $h$ where it did not (H2). The three tasks show why.

\textbf{H1, diamond at 2.0\,eV: the verifier makes the difference.} H1 was the hardest task, and its failures share one cause. The HSE-scissor gap used during search is calibrated at zero strain and overestimates the strained diamond gap, by 0.12--0.57\,eV at the agents' final candidates. Every arm that trusted it without a full-HSE06 check missed the $\pm$0.03\,eV window: all cBO-expert and cBO-agent states fell at 1.82--1.96\,eV (with the corrected two-sided encoding; Supplementary Table~\ref{tab:S2}), and the cBO-naive arm, which thresholded the raw PBE gap, landed at 2.97--3.03\,eV. The verifier rejected five H1 candidates. One agent run re-targeted after its rejection and passed at 1.984\,eV (\hval{30.334}); with the $\Delta$ correction at the valley minimum it lies at 1.962\,eV, just below the window (Section~\ref{sec:absgap}). The other two ended at 1.938 and 1.502\,eV. The scripted arm passed on two of three seeds in the same way, through its scripted re-target, and its 29.103\,\hunit{} state is the H1 reference.

\textbf{H2, Si gap below the 1550\,nm photon energy: the formalization matters.} All four agent runs were verified, with full-HSE06 gaps of 0.727--0.773\,eV, and their states are elastically stable. The lowest-$h$ state came from the cBO-naive arm (\hval{0.281}, $\EgL{HSE06} = 0.789$\,eV), but by chance: its raw-PBE constraint, a gap of at most 0.80\,eV, never binds for Si, whose PBE gap stays below 0.80\,eV at every strain computed, so it minimized $h$ with no effective gap constraint and would have accepted unstrained Si. The scripted arm given the same naive formalization shows this: its first candidate was nearly unstrained Si (\hval{0.007}), which the full-HSE06 check rejected at 1.171\,eV, and the scripted arm passed only after re-targeting (Table~\ref{tab:outcomes}; Supplementary Section~\ref{sec:S10}). The agents instead kept a margin and targeted 0.78\,eV (Fig.~\ref{fig:h2}a); against the naive reference their regrets are 0.008--0.070\,\hunit{} (Fig.~\ref{fig:h2}b).

\textbf{H3, metallic Si in the $\varepsilon_{33} = 0$ subspace.} All three agent runs and all three scripted runs were verified, and every cBO-expert and cBO-agent run was metallic at full HSE06. The cBO-agent run coincides with cBO-expert seed 0 (2.698\,\hunit) because the two formalizations reduce to the same constraint and bounds, and two agent runs reached the same state (2.451\,\hunit) because they share the optimizer's default seed (Section~\ref{sec:limits}). The cBO-naive run, added after the data freeze, failed: searching the full 6D strain space with a raw-PBE gap of at most 0\,eV, it found a state (\hval{1.009}) that is not metallic at full HSE06 (0.414\,eV) and has $\varepsilon_{33} = 2.3\%$, outside the H3 constraint.

\textbf{The agent runs its own active learning.} In earlier deep-ESE campaigns an expert chose the active-learning schedule (Section~\ref{sec:intro}); here the agent did. Nine of the 14 designer runs closed the loop at least once (fit, optimize, compute DFT at the new states, refit), fitting the surrogate 2--6 times. Six of them also called the active-learning tool and treated its proposals as advice rather than orders; \runid{H2_s2}, for example, asked for 8 points but computed 1, exploiting the optimizer's candidate. The five runs that fitted once (two H2 and all three H3 runs) were still verified: on these tasks one round of surrogate-guided search was enough.

\subsection{Verifier fault injection and formalization accuracy}\label{sec:fault}

\textbf{The verifier catches planted errors.} Is each agent competent on its own? We took the verifier inputs of three H2 runs, planted one fault in each copy and ran the verifier single-turn with tools off, three copies per fault type (Fig.~\ref{fig:outcomes}b). The faults are explicit contradictions in the evidence (a PBE gap reported as HSE06, a gap violating the target, misreported $h$, an unconverged SCF, an elastically unstable state, a misreported strain), so the test measures whether the verifier cross-checks the evidence it is given, not whether it detects subtle physics errors. It rejected all 18 transcripts carrying these six faults and verified all three clean controls. It missed a seventh fault, strain given in percent instead of fractions (0 of 3), a slip the tool schema stops upstream: the schema rejects any strain component above 0.25, so a percent-unit input passes only if every intended component is below 0.25\%. In the ten live held-out runs the verifier examined 13 candidates and rejected 5, all on H1 for scissor error.

\textbf{The designer writes correct problems.} Claude Sonnet formalized 22 paraphrased goals single-turn with tools off, and two automated judges scored the blinded answers against the pre-registered rubric (Fig.~\ref{fig:outcomes}c; dated amendment). Mean scores were 0.97 and 0.90 of the maximum. Both judges passed every answer on minimizing elastic energy, evaluating thresholds at a corrected level of theory and including stability where the goal asked for it. Most disagreements concerned a safety margin on the right constraint, such as $\Eg \le -0.05$\,eV for metallization or a 0.745\,eV screen for 1550\,nm, which one judge accepted and the other did not; one H1 answer wrote a one-sided constraint, $\Eg \le 2.0$\,eV, which both marked wrong, and the subspace misses were on H3. The judges are Claude models, like the agent; a deterministic rule-based scoring of the same answers against the same key, with no model involved, gave a mean of 0.945 and agreed with the two judges on 94\% and 90\% of scores, but neither replaces independent expert judgment. With five underlying goals and no human raters, this is a small demonstration, not a validation of expert-free problem formulation.

\subsection{Physics of the reported states}\label{sec:physics}

\begin{figure}[!tbp]
  \centering
  \includegraphics[width=\textwidth]{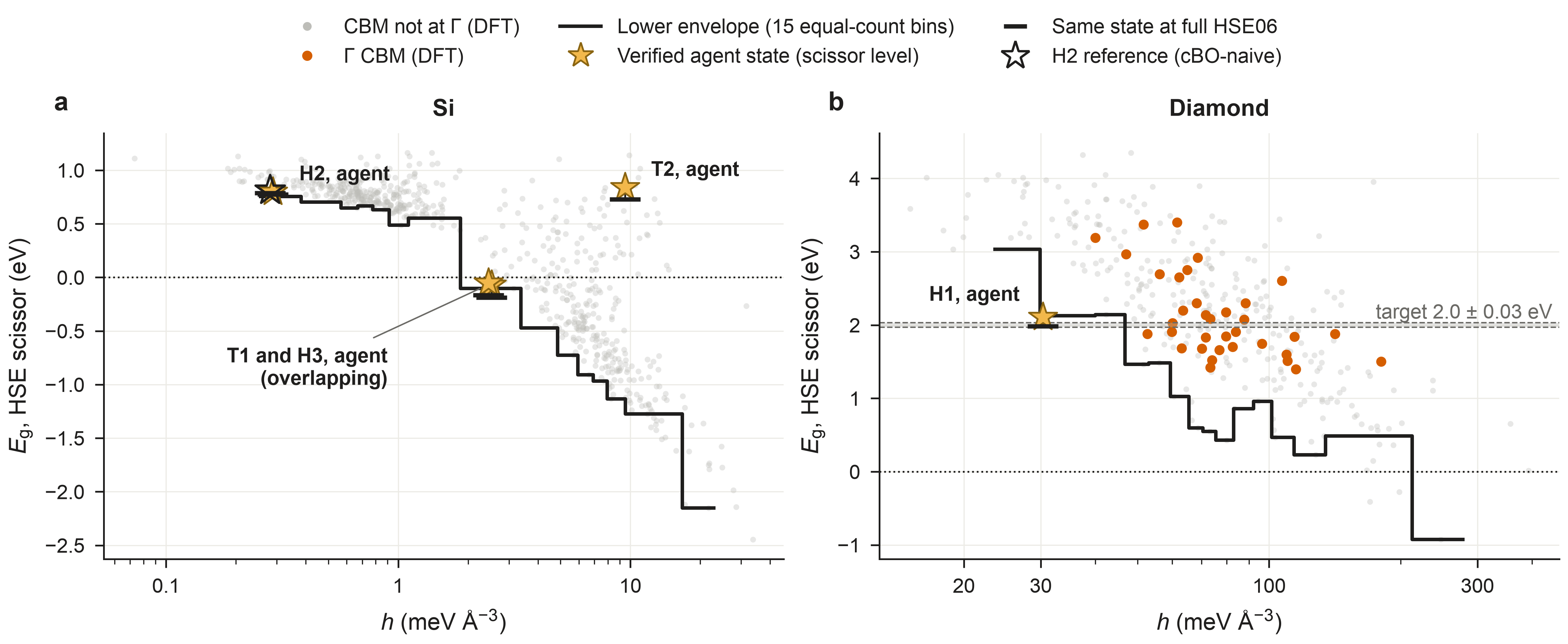}
  \caption{\textbf{Verified states on the gap--energy maps of Si and diamond.} HSE-scissor gap (eV) vs elastic energy density $h$ (\hunit; log $x$ axis) for the DFT benchmark sets of (a)~Si and (b)~diamond. The lower envelope is the lowest gap in each of 15 equal-count bins in $h$, drawn as a step centered on each bin; vermilion points have a $\Gamma$ conduction-band minimum. Stars mark each task's lowest-$h$ verified agent state at the same scissor level as the map; the bar under each star is the same state at full HSE06 (valley-sampled; Section~\ref{sec:absgap}). The open star is the H2 reference from the cBO-naive arm. In (a) the T1 and H3 states overlap and share one label. In (b) the $2.0 \pm 0.03$\,eV target band is shaded.}
  \label{fig:maps}
\end{figure}

Are the verified states physically interesting? Fig.~\ref{fig:maps} places the lowest-$h$ verified agent state of each task on the Si and diamond gap--energy maps; each was accepted by the verifier at full HSE06 and is elastically stable.

\textbf{Si metallization (T1).} Both T1 runs were verified, at $h = 2.530$ and 2.558\,\hunit{} (full-HSE06 gaps $-0.188$ and $-0.208$\,eV), above the published 6D minimum of 1.35\,\hunit~\cite{shi2019}. Part of the difference is the level of theory: at our level the published state, as corrected in 2020, needs \hval{1.439} and is metallic at PBE but not at full HSE06 (gap +0.246\,eV), so a state that must close the gap at full HSE06 needs more elastic energy. The rest is a search shortfall: the lowest H3 states, confined to a subspace of the same 6D strain space, need only 2.331 (scripted) and 2.451\,\hunit{} (agent), so the T1 runs also stopped above the 6D optimum at our level.

\begin{figure}[!tbp]
  \centering
  \includegraphics[width=\textwidth]{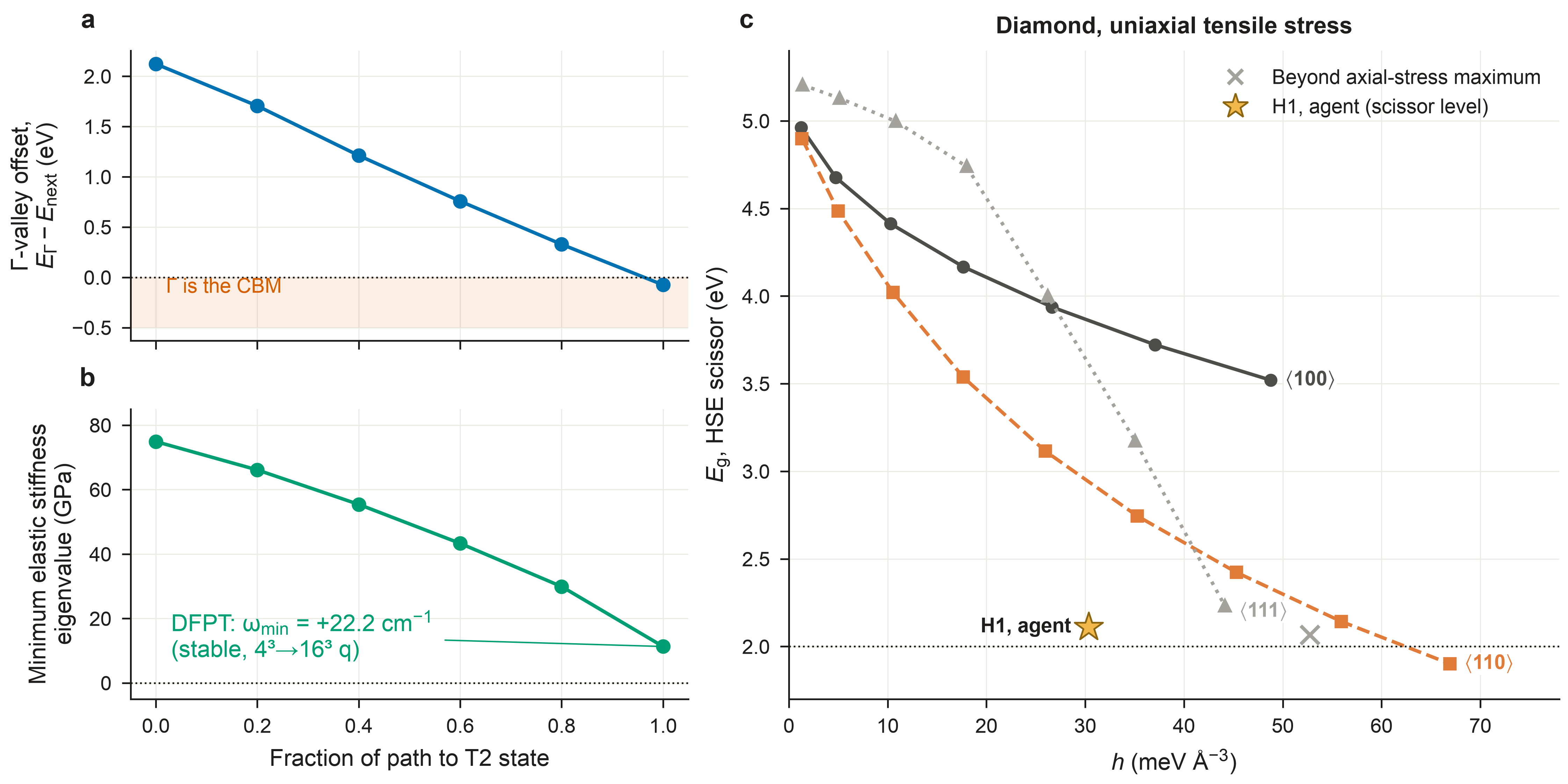}
  \caption{\textbf{Stability along the T2 path and diamond uniaxial routes.} (a)~$\Gamma$-valley offset, the energy of the $\Gamma$ valley minus the next-lowest conduction valley (HSE scissor), along the straight strain path from zero to the agent's direct-gap Si state (T2); below zero $\Gamma$ is the conduction-band minimum. (b)~Lowest eigenvalue of the finite-strain elastic stiffness (second Piola--Kirchhoff stress vs Green--Lagrange strain, Section~\ref{sec:stability}) along the same path; DFPT at the endpoint gives $\wmin = +22.2$\,\cmi. (c)~Diamond gap (HSE scissor, eV) vs $h$ (\hunit) under uniaxial tensile stress along \dir{100}, \dir{110} and \dir{111} (residual transverse stress removed), with the agent's H1 state at the same level; the dotted line marks 2.0\,eV; $\times$ marks the \dir{111} state beyond the maximum axial stress, where uniaxial loading is unstable.}
  \label{fig:physics}
\end{figure}

\textbf{A stable route to direct-gap Si (T2).} The agent reached a direct gap at $\Gamma$ at \hval{9.476}, with $\EgL{HSE06} = 0.727$\,eV and the $\Gamma$ valley 0.174\,eV below the next conduction valley ($\Delta$) at full HSE06. The principal strains of this state, 15.1\%, 1.0\% and 0.0\%, amount to a nearly uniaxial strain of about 15\% (Green--Lagrange; 14.1\% engineering) along an axis close to \dir{111}. This is close to the largest elastic strain measured in Si nanowires (about 16\%~\cite{zhang2016}), but a nanowire under uniaxial stress also contracts laterally, so the agent's state is not the nanowire state. The triaxial 9.3\% state reported in 2019~\cite{shi2019} needs \hval{14.672} at our level of theory (15.801\,\hunit{} if its strain is read as linear), so the agent's state needs 35--40\% less elastic energy; the 2019 envelope used GW-fused gaps, so this comparison does not show it to be wrong. Along the straight path to the agent's state, $\Gamma$ becomes the conduction-band minimum between 80\% and 100\% of the path (Fig.~\ref{fig:physics}a), and the lowest elastic stiffness eigenvalue softens from 74.9 to 11.3\,GPa but stays positive (Fig.~\ref{fig:physics}b). DFPT at the endpoint gives no imaginary mode (lowest reported phonon frequency $\wmin = +22.2$\,\cmi).

\textbf{Si gap below the 1550\,nm photon energy (H2).} The H2 states, the agents' and the reference alike, are close to uniaxial compression, $-2.4$ to $-2.9\%$ along one cubic axis, at $h \approx 0.3$\,\hunit: compression lowers the $\Delta$ valley along that axis until the HSE06 gap falls below 0.80\,eV. DFPT at the Opus run's state gives no imaginary mode ($\wmin = +26.6$\,\cmi). The gap stays indirect, as the designer itself noted (Fig.~\ref{fig:h2}a), so absorption at 1550\,nm would be phonon-assisted and weak; the 0.80\,eV threshold is only a proxy for absorption. Strain-extended absorption to 1550\,nm has been measured in Si nanomembranes under biaxial tension of up to about 3.5\%, rather than the near-uniaxial compression found here~\cite{katiyar2020}.

\textbf{Diamond at 2\,eV versus uniaxial loading (H1).} The H1 task also asked for a comparison with uniaxial loading, which we computed along the three high-symmetry directions with the residual transverse stress removed (Fig.~\ref{fig:physics}c, HSE-scissor level). \dir{100} tension reaches only 3.52\,eV at 13.8\% axial strain (axial strains in this paragraph are Green--Lagrange). \dir{111} reaches its maximum axial stress, 82.7\,GPa, near 14\% strain at a gap of 2.23\,eV and is unstable beyond the stress maximum, so it cannot reach 2\,eV as a stable uniaxial state. \dir{110}, the only uniaxial route that reaches 2\,eV within the computed range, reaches the agent's gap at about \hval{57.4} (interpolated), both at the scissor level (2.11\,eV) and at full HSE06 (1.984\,eV): about twice the agent's 30.3\,\hunit. The agent's state, with principal strains of $-7.6\%$, $-0.5\%$ and $+6.9\%$, is phonon-stable ($\wmin = +96.9$\,\cmi{} at relaxed atomic positions), but no experiment has yet produced homogeneous diamond with 7\% compression and 7\% tension, so it is a target for multiaxial loading rather than a demonstrated state. At about 10\% axial strain, close to the uniform tension demonstrated in microfabricated diamond~\cite{dang2021}, stresses along \dir{100} and \dir{110} give gaps of only 3.94 and 3.11\,eV here. Reaching 2\,eV therefore needs either much larger uniaxial strain, about 16\% along \dir{110}, or a shear-containing multiaxial state such as the agent's.

\section{Discussion}\label{sec:discussion}

\subsection{What the language model adds}\label{sec:llm}

On these tasks the scripted arm, an LLM-free ablation of the same toolkit, did as well as the agent, so reliability came from the toolkit and verification loop, and the language model's search decisions added nothing measurable. What the language model contributes is the step the script was given: turning a one-sentence goal into a correct formalization (Section~\ref{sec:fault}), with its own choice of margin, subspace and level of theory, without an expert. Given to cBO, the agent's formalization did no better than the expert's (Fig.~\ref{fig:outcomes}a); both are correct where the naive one is not (Section~\ref{sec:campaigns}). On H1 the agent and scripted arms passed only after the full-HSE06 check exposed the scissor error and they re-targeted, which cBO, with no higher-fidelity check and a smaller budget, could not do. The comparison therefore shows the value of the verification loop rather than ranking the optimizers.

Unlike agents that execute first-principles workflows~\cite{liu2025vaspilot,harikumar2026,yang2026quasar}, this work performs inverse design over a continuous strain tensor, assigns the acceptance tools to an independent verifier and measures reliability against an LLM-free ablation. Relative to our earlier deep-ESE studies~\cite{shi2019,tsymbalov2021,shi2024,shi2020}, which relied on expert-run campaigns, the new elements are the agentic formalization and verification layer, the valley-resolved equivariant surrogate and an evaluation on pre-registered tasks. The reported states are also new: a multiaxial diamond state at 2\,eV and a phonon-stable, nearly uniaxial route to direct-gap Si that needs 35--40\% less elastic energy than the published triaxial state at the same level of theory (Section~\ref{sec:physics}).

The independent verifier and the LLM-free ablation, together with an evaluation on pre-registered tasks in which every reported number was checked against the results file, should carry over to other agentic design problems; without the ablation, the agent's results could easily have been credited to the language model.

\subsection{Limitations}\label{sec:limits}

Run counts are small: three agent runs per held-out task (four on H2), three scripted seeds per task, five cBO-expert seeds on H1 and H2, two cBO-expert seeds on H3 and one run per task for cBO-agent, cBO-naive and the scripted arm with the naive formalization; the scripted arm and the two-sided-window H1 cBO rerun were defined after the agent outcomes were known, and the H3 cBO-naive run and the scripted naive runs were added after the data freeze, when all other results were known. References are the best verified states found, not proven optima. Formalization accuracy was scored by automated judges from the agent's model family, not by human experts. Thirteen of the 14 agent runs called the sampling and optimization tools with their default random seed, so replicates of a task started from the same initial samples (\runid{H3_s1} and \runid{H3_s2} reached the same state), and the run-to-run spread is understated. cBO also differs from the toolkit in its gap representation, a Gaussian process on \Eg{} rather than valley energies, so its outcomes do not separate the search algorithm from the representation. All results are at the PBE/HSE06 level without spin--orbit coupling or temperature. Comparing agents fairly will need a shared benchmark with held-out goals, frozen baselines and independent human rating. The tasks, harness and transcripts, to be released with this paper, could form part of one.

\subsection{Outlook}\label{sec:outlook}

This study covered two diamond-structure crystals under homogeneous strain, with the band gap as the only target property. Four extensions look most useful to us.

The first is to let the agent choose the level of theory state by state. The zero-strain HSE scissor overestimated the full-HSE06 gap at every final agent candidate, by 0.02--0.12\,eV in Si, where no candidate was rejected for it, and by 0.12--0.57\,eV in strained diamond, where it caused all five H1 rejections (Section~\ref{sec:campaigns}). GW-fused surrogates for Si and diamond already exist~\cite{shi2019,tsymbalov2021}. An agent could decide when a candidate sits close enough to a decision boundary to justify a hybrid or GW calculation, and pay for that fidelity only there.

The second is to bring stability into the search instead of checking it afterwards. Here the designer had no stability tool: elastic stability was checked only by the verifier, DFPT was run on three reported states, and the \dir{111} diamond route lost stability before reaching 2\,eV (Section~\ref{sec:physics}). The 6D phonon-stability boundary of diamond has already been learned~\cite{shi2024}; offered to the agent as a tool, it would keep the search inside admissible strains.

The third concerns what experiments can reach. Real samples carry strain gradients, and a given geometry can produce only some strain states~\cite{banerjee2018,dang2021}. The constrained task H3 is a small step in this direction; a traction-free film, in which the stresses normal to the surface vanish rather than the normal strain, is the experimentally relevant case and is left for future work. Coupling the agent to a finite-element model of the loading would turn a target property into a loading protocol. Other properties fit the same framework: effective masses, band alignments, lattice thermal conductivity, whose strain dependence has been mapped for diamond and Si~\cite{shi2024,shi2026eml}, and optical absorption, beyond the gap proxy used for H2. Moving beyond group IV mainly requires tracking valleys by band character in lower-symmetry crystals.

The fourth is to make the verifier itself more robust. The percent-unit fault it missed (Section~\ref{sec:fault}) was caught only by the strain bounds of the tool layer. Unit and plausibility checks, recomputation of the reported $h$, gap and stress from the strain, and consistency checks between them belong in deterministic code that runs before the verifier's judgment, with a broader suite of subtle injected faults (sign flips, swapped Voigt components, the wrong strain measure) kept as a regression test.

\section{Conclusions}\label{sec:conclusions}

An LLM agent can carry out inverse design in deep ESE: starting from one plain-language sentence, it reached DFT-verified, stability-checked strain states in Si and diamond. Three design choices made this possible. The language model posed the problem (objective, constraints, subspace, margin and level of theory) but left the numerical search to a deterministic optimizer. The optimizer ran on a surrogate of smooth valley energies rather than of the gap itself; this surrogate was accurate enough with tens of DFT points to guide a search on a small budget, though not to certify a state. And full HSE06 and elastic stability were reserved for an independent verifier, so the designer could not certify its own answer; on H1 the verifier rejected five candidates whose full-HSE06 gaps missed the target.

On the held-out tasks the agent and the scripted, LLM-free ablation did about equally well, with 8 of 10 and 8 of 9 runs verified. Bayesian optimization, which searched at the scissor or raw-PBE level without a higher-fidelity check, never met the diamond 2\,eV target at full HSE06. On this evidence, within the limits of Section~\ref{sec:limits}, the language model's contribution is the formalization, not its search decisions. DFPT confirmed the phonon stability of the direct-gap (T2) Si state, of one Si state for the 1550\,nm task (H2) and of the diamond 2\,eV state. The framework, the pre-registration and every transcript will be released with the paper.

\section*{Declarations}

\textbf{Authorship contribution statement.} Zhe Shi: Conceptualization, Methodology, Investigation, Formal analysis, Writing -- original draft, Writing -- review \& editing.

\textbf{Declaration of competing interest.} The author declares that they have no known competing financial interests or personal relationships that could have appeared to influence the work reported in this paper.

\textbf{Funding.} This research did not receive any specific grant from funding agencies in the public, commercial, or not-for-profit sectors.

\textbf{Declaration of generative AI usage.} Large language models are part of the research method: the designer and verifier agents are Claude models from Anthropic, run headless through Claude Code. Their configuration and model identifiers are given in Methods and the Supplementary Information. The author also used Claude to revise text of the manuscript and Supplementary Information from the author's results and instructions, to check the text against the results file and to polish the English. After using this tool, the author reviewed and edited the content as needed and takes full responsibility for the content of the publication.

\textbf{Data and code availability.} The data and code that support the findings of this study (the toolkit, MCP server, agent harness, pre-registration and amendments, reference datasets, agent transcripts, scoring scripts and the source data for every figure) will be deposited in a public repository with a DOI on acceptance; until then they are available from the author on request.

\textbf{Acknowledgments.} The author thanks Ju Li, Subra Suresh and Ming Dao for years of collaboration on deep elastic strain engineering.


\clearpage
\setcounter{section}{0}
\setcounter{table}{0}
\setcounter{figure}{0}
\renewcommand{\thesection}{S\arabic{section}}
\renewcommand{\thetable}{S\arabic{table}}
\renewcommand{\thefigure}{S\arabic{figure}}
\renewcommand{\theHsection}{S\arabic{section}}\renewcommand{\theHtable}{S\arabic{table}}\renewcommand{\theHfigure}{S\arabic{figure}}

\phantomsection
\addcontentsline{toc}{section}{Supplementary Information}
\begin{center}
  {\Large\bfseries Supplementary Information}\\[0.6em]
  {\large Agentic Deep Elastic Strain Engineering of Semiconductor Electronic Properties}\\[0.4em]
  Zhe Shi\\
  {\small Department of Materials Science and Engineering, Massachusetts Institute of Technology, Cambridge, MA 02139, USA}\\
  {\small \texttt{shizhe1@mit.edu}}
\end{center}
\bigskip

\section{System prompts and tool access}\label{sec:S1}

All reported runs used the designer--verifier harness of main-text Section~\ref{sec:framework} (Claude Code 2.1.305, headless). The prompts below are verbatim from \path{eseagent/scripts/run_agent.py}, unchanged in every reported run; British spellings are as in the code.

\textbf{Designer, system prompt.} ``You are the inverse-design specialist. First write down the optimisation problem precisely: objective, every constraint with its numerical threshold and the level of theory at which it must be evaluated, the strain subspace and bounds, and any stability requirement. Then run an efficient campaign with the ese tools (sample -> DFT -> fit surrogate -> active learning -> optimise -> DFT check), staying within budget. The numerical search is done by optimize\_strain; you choose its inputs. Report candidates with their DFT values and your reasoning in terms of conduction-band valleys.\newline
Finish with the requested JSON: your formalisation and your single best DFT-checked candidate.''

\textbf{Designer, user prompt.} The task sentence of main-text Table~\ref{tab:tasks}, verbatim. After a rejection, the second designer session receives the same sentence followed by: ``An independent verifier REJECTED your previous candidate [strain] for these reasons: [the verifier's issues]. Revise and try again.''

\textbf{Verifier, system prompt.} ``You are an independent verifier. You did not design the candidate. Re-evaluate it with DFT, check symmetry-equivalent strains, units and conventions (fractions, tensor shear, Green-Lagrange), whether every constraint really holds at the right level of theory, and stability where requested. Reject any claim that lacks DFT evidence. Report clearly: VERIFIED or REJECTED, with numbers.''

\textbf{Verifier, user prompt.} ``Task given to the design team: [task sentence]\newline
Their formalisation: [the designer's formalization, JSON]\newline
Their candidate strain (Green-Lagrange, Voigt, tensor shear, fractions): [strain]\newline
Their claimed h=[h] meV/A\textasciicircum 3, gap=[gap] eV. Verify independently.''

\textbf{Tool access.} Designer: \tool{describe_material}, \tool{sample_strains}, \tool{run_dft}, \tool{fit_surrogate}, \tool{query_surrogate}, \tool{optimize_strain}, \tool{propose_active_learning}, \tool{map_envelope}, \tool{log_finding}. Verifier: \tool{describe_material}, \tool{run_dft}, \tool{verify_hse}, \tool{elastic_stability}, \tool{log_finding}. Both sessions see the full ese tool list, but calls outside a session's allow-list are refused by Claude Code; the designer transcripts record such refusals for \tool{verify_hse} and \tool{elastic_stability}. Neither session has file, shell or web access. Each session returns JSON against a fixed output schema (formalization, strain, $h$, gap, CBM valley and whether the constraints are met for the designer; verdict, strain, $h$, gap, level of theory, stability and issues for the verifier). The fault-injection replays (main-text Section~\ref{sec:fault}) used the verifier system prompt.

\section{Exploratory surrogate study (dated amendment)}\label{sec:S2}

This study was added after the campaigns as a dated amendment to the pre-registration. All agent and scripted campaigns in main-text Section~\ref{sec:campaigns} used the original model (main-text Table~\ref{tab:mae}, first column), and every reported state was verified by DFT, so the results below change no outcome. Models were selected on a validation split and scored on the same five test splits as main-text Table~\ref{tab:mae}.

\textbf{Scaling.} Beyond a few hundred training points the valley-resolved error levels off while the direct-regression baseline keeps improving: for diamond, 112.5\,meV at 500 points and 103.7\,meV at 6,456 (direct regression 279.5\,$\rightarrow$\,170.7\,meV; Table~\ref{tab:S4}); for Si over the wide range, 34.6\,meV at 500 and 34.8\,meV at 1,018 (Table~\ref{tab:S7}). We therefore tested model changes rather than more data. A later extension of the diamond training set to 6,456 states confirmed that more data does not close the accuracy gap (Section~\ref{sec:S9}, Table~\ref{tab:S4}).

\begin{table}[!htbp]
  \centering
  \caption{Held-out gap MAE (meV), mean $\pm$ s.d.\ over the five test splits of main-text Table~\ref{tab:mae}: original valley-resolved model and the larger, longer-trained network (``improved'').}
  \label{tab:S1}
  \small
  \begin{tabular}{@{}lcc@{}}
    \toprule
    \textbf{Benchmark set} & \textbf{Valley-resolved (campaigns)} & \textbf{Valley-resolved, improved} \\
    \midrule
    Si, small strain ($|\varepsilon_{ii}| \le 3\%$) & 9.2 $\pm$ 0.7 & 3.8 $\pm$ 0.6 \\
    Si, wide range ($|\varepsilon_{ii}| \le 10\%$) & 38.5 $\pm$ 4.8 & 24.6 $\pm$ 4.5 \\
    Diamond ($|\varepsilon_{ii}| \le 10\%$) & 123.6 $\pm$ 24.3 & 77.6 $\pm$ 21.3 \\
    \bottomrule
  \end{tabular}
\end{table}

The improved network is a larger, longer-trained network from the exploratory study, selected on validation data and tested on the same five splits.

\begin{itemize}[leftmargin=*]
  \item \textbf{A larger, longer-trained network} (width 128, three or four hidden layers, 1,500--3,000 epochs) lowered the full-data error to 3.8 $\pm$ 0.6\,meV for Si at small strain, 24.6 $\pm$ 4.5\,meV for Si over a wide range and 77.6 $\pm$ 21.3\,meV for diamond: 1.6--2.4$\times$ better than the networks used in the campaigns. At 50 points the gain was smaller, and for wide-range Si there was none.
  \item \textbf{An explicit Bir--Pikus cone term} (main-text ref.~\cite{bir1974}) for the degenerate valence-band maximum, with learnable deformation potentials, did not help. At full training size it was worse than the larger network alone on every set (7.3, 27.8 and 99.0\,meV); at 50 points it was within the split-to-split spread.
\end{itemize}

\textbf{Further attempts.} A symmetry-invariant Gaussian process per valley family was worse than the larger network on every set and size (5.6, 28.1 and 150.8\,meV at full data). Wider and deeper networks (width 256, four hidden layers) lowered the validation error by 10--18\% across the three sets, while longer training and a ten-member ensemble gave little; on the five diamond test splits the widest network gave 77.7 $\pm$ 19.8\,meV, no better than 77.6 $\pm$ 21.3\,meV, so the validation gains did not carry over. Adding 300 diamond states chosen where the ensemble was most uncertain more than doubled the diamond training data but left the test error unchanged (78.5 $\pm$ 18.1\,meV against 77.6 $\pm$ 21.3\,meV). As a control, 300 random states from the benchmark distribution gave 72.9 $\pm$ 25.5\,meV, lower on four of the five splits but within the split-to-split spread. With 800 added random states, which bring the diamond training data to about four times its original size, the error was 71.3 $\pm$ 20.4\,meV: an 8\% gain, again within the spread, with one test split getting worse. The uncertainty-targeted states, concentrated at extreme strain, did worse than random ones.

\textbf{Error analysis.} An analysis of the largest test errors found no faulty DFT labels. Most of the remaining error comes from two situations the valley representation cannot express: states where the valence-band maximum leaves $\Gamma$ under extreme strain, and diamond states whose conduction-band minimum is the degenerate $\Gamma$ valley. An order-of-magnitude gain would need a representation that also tracks the valence band away from $\Gamma$ and the $\Gamma$ conduction states, not more tuning.

Sources: \path{results.json} keys \path{surrogate_study}, \path{gp_study}, \path{surrogate_step3_val}, \path{diamond_A_test}, \path{al_diamond}, \path{rand_diamond}, \path{rand800_diamond}; scripts \path{eseagent/scripts/surrogate_study.py}, \path{gp_study.py}, \path{diamond_B.py}, \path{al_diamond.py}.

\section{Formalization rubric and the judges' answer key}\label{sec:S3}

Verbatim from \path{eseagent/scripts/rate_formalization.py}. Each judge (Claude Sonnet, Claude Haiku; single turn, tools off) received the rubric as its system prompt and, as the user prompt, ``Answer key: [key for the task]'', the goal given to the model and the model's answer. Criterion R4 is not scored (not applicable) for goals that do not ask for stability (H1, H3, T1); the scored answers are in \path{rating/rating_sheet.csv}.

\textbf{Rubric (system prompt).} ``Score one formalisation answer on five criteria, each 0 or 1:\newline
R1: the objective is to minimise elastic strain energy (energy density h or total elastic energy).\newline
R2: the constraint has the correct type and threshold given the answer key (e.g.\ 1550 nm -> Eg <= 0.80 eV).\newline
R3: absolute gap thresholds are to be evaluated at a corrected level of theory (HSE06/hybrid, GW, or experiment-anchored), not raw PBE/LDA. For a relative criterion (direct gap) score 1 if the answer evaluates the valley ordering at a corrected level or explicitly addresses the level-of-theory dependence.\newline
R4: a stability requirement (elastic/Born and/or phonon) is included as a constraint or required check.\newline
R5: the strain subspace matches the key (e.g.\ e33 = 0 for the film task) and strain units/conventions are stated or unambiguous.\newline
Score strictly against the key. Give one short reason per criterion.''

\textbf{Answer key.}
\begin{itemize}[leftmargin=*]
  \item H1: ``diamond; target gap Eg = 2.0 eV (an equality with a stated tolerance, or a two-sided window around 2.0 eV); full 6D strain.''
  \item H2: ``silicon; Eg <= 0.80 eV (1240/1550 nm); full 6D strain; stability required.''
  \item H3: ``silicon (001) film; Eg <= 0 (metallic); strain subspace with e33 = 0 (other five components free).''
  \item T1: ``silicon; Eg <= 0 (metallic); full 6D strain.''
  \item T2: ``silicon; direct gap: the Gamma conduction-band valley lies below every other conduction valley (Delta, L); full 6D strain; stability required.''
\end{itemize}

The answer key was written when the automated judges replaced the human raters (dated amendment of 2026-10-24; Table~\ref{tab:S2}), after the formalization answers had been generated.

\section{Transcripts}\label{sec:S4}

Full designer and verifier transcripts for every run (\path{transcript_*.jsonl}), with each run's ledger, data store and final verdict, and the scripted-arm logs will be deposited in a public repository with a DOI on acceptance; until then they are available from the author on request.

\section{Pre-registration and amendments}\label{sec:S5}

The full pre-registration with its dated amendments (\path{preregistration.md}) will be deposited in the same repository on acceptance and is available from the author on request until then; Table~\ref{tab:S2} summarizes the deviations.

{\footnotesize
\setlength{\LTcapwidth}{\textwidth}
\begin{longtable}{@{}>{\raggedright\arraybackslash}p{0.25\textwidth} >{\raggedright\arraybackslash}p{0.47\textwidth} >{\raggedright\arraybackslash}p{0.22\textwidth}@{}}
  \caption{Deviations from the pre-registration.}\label{tab:S2}\\
  \toprule
  \textbf{Pre-registered} & \textbf{What was done} & \textbf{Reason} \\
  \midrule
  \endfirsthead
  \multicolumn{3}{@{}l}{\small\itshape Table~\ref{tab:S2} (continued)}\\
  \toprule
  \textbf{Pre-registered} & \textbf{What was done} & \textbf{Reason} \\
  \midrule
  \endhead
  \bottomrule
  \endlastfoot
  \kgrid{10} $k$-mesh & \kgrid{8} for production runs & cost; amended before any agent run \\
  \addlinespace
  Frozen oracle from non-agent arms ($\ge 10$ cBO runs plus local refinement) & Reference = best successful state from any arm, agent and scripted-arm states included & compute; reduced scope \\
  \addlinespace
  cBO at 150 units, $\ge 10$ runs & 100 units: cBO-expert 5 seeds (H3: 2), cBO-naive 1 (the H3 run added on 2026-10-28, after the data freeze; dated amendment), cBO-agent 1 & compute \\
  \addlinespace
  Expert formalization encoded as written & First H1 cBO runs encoded ``hitting 2~eV'' one-sided ($\Eg \le 2.0$\,eV); rerun with the two-sided window, both reported & encoding error found in internal review \\
  \addlinespace
  Two blinded human raters, Cohen's $\kappa$ & Two automated judges (Claude Sonnet and Haiku); $\kappa$ between them & no human raters available in time \\
  \addlinespace
  Faults: unit slip, engineering vs tensor shear, valley mislabel, unconverged flag, hidden constraint violation & Replayed transcripts with seven fault types (six explicit contradictions and a percent-unit slip) plus clean controls. A first design that replayed only scissor gaps rejected every item, clean controls included, and was replaced & first design invalid \\
  \addlinespace
  Agent arm (harness not specified) & A deterministic harness runs the designer and the verifier as separate sessions with role-specific tool sets and one feedback round (dated amendment of 2026-10-23, made after two H2 runs had started under an orchestrator agent with sub-agents). Those two runs were aborted by a harness bug, are kept with the run records and are not counted; \runid{H2_s1} and \runid{H2_s2} were then run with the new harness & harness bug: in headless mode the orchestrator returned before the designer had finished \\
  \addlinespace
  Absolute gap targets evaluated with an HSE correction & After the H1 agent runs, every arm's H1 state was scored at full HSE06; for cBO the five lowest-$h$ scissor-feasible states of each run were re-checked and the best passing one reported (dated amendment of 2026-10-23) & drift of the zero-strain scissor in strained diamond; the success criterion was applied, not changed \\
  \addlinespace
  Main-text Fig.~\ref{fig:h2}a: the H2 run with the median regret & Median over the three Sonnet replicates, which selects \runid{H2_s3} (dated amendment of 2026-10-24) & four runs have no single middle run; ranking by $h$ equals ranking by regret against any reference \\
  \addlinespace
  Permuted-feedback run per held-out task & Not run & compute and usage limits \\
  \addlinespace
  Primary endpoint: relative regret at 150 units; success needs $h \le 1.10\,h_\mathrm{oracle} + 0.05$\,\hunit & Absolute regret against the best state found & no frozen oracle \\
  \addlinespace
  Full HSE06 re-evaluation & Valley-sampled HSE06 (main-text Section~\ref{sec:absgap}) & cost \\
  \addlinespace
  --- & Added after the outcomes were seen (exploratory): scripted arm; H1 cBO rerun with the two-sided window; repeated surrogate splits & requested by internal review \\
  \addlinespace
  --- & Added after the outcomes were seen, as dated amendments (2026-10-24 to 2026-10-26): larger surrogate networks and targeted and random diamond training data (Section~\ref{sec:S2}); full-HSE06 check of \dir{110} uniaxial diamond, used for the \dir{110} comparison of main-text Section~\ref{sec:physics}. Without an amendment: phonon-stability classifier cross-check (Section~\ref{sec:S7}) and offline replay with the larger surrogate (Section~\ref{sec:S8}) & exploratory; no success count or reference changed \\
  \addlinespace
  --- & Added after the data freeze, when all other results were known (dated amendments of 2026-10-28; the H3 cBO-naive run is listed above): scripted arm with the naive formalization, $\Delta$-line check of the full-HSE06 gap and rule-based scoring of the formalizations (Section~\ref{sec:S10}); extended diamond learning curve (Table~\ref{tab:S4}); extended Si learning curves (Table~\ref{tab:S7}, 2026-10-30) & exploratory; no campaign outcome or reference changed \\
  \addlinespace
  Elastic stability from DFT stresses & Campaign tool used Cauchy stress; all 35 states with a full stencil re-evaluated afterwards with the second Piola--Kirchhoff stress (no verdict changed) & correct finite-strain measure (post-campaign correction) \\
  \addlinespace
  --- & Added after external review (2026-10-31): $\Delta$ correction at the valley minimum and H1 DFPT at relaxed internal coordinates (Section~\ref{sec:S10}, Tables~\ref{tab:S8} and~\ref{tab:S9}) & review check; scoring, success counts and references unchanged \\
\end{longtable}
}

\section{Wall-clock time and tokens per agent run}\label{sec:S6}

Logged values only: wall-clock time from each run's \path{meta.json} (start to end, designer plus verifier, including DFT) and tokens summed over the result records of the designer and verifier transcripts. \runid{H3_s2} followed the same trajectory as \runid{H3_s1} (shared optimizer seed); its DFT calls were served from the content-addressed cache, which charges a hit like a miss: the run's ledger records the start of \tool{verify_hse} (about 30\,min when uncached) and of \tool{elastic_stability} (12 DFT calculations) less than 7\,s before the run ended.

\begin{center}
\small
\begin{tabular}{@{}lrrrrr@{}}
  \toprule
  \textbf{Run} & \textbf{Wall-clock (min)} & \textbf{Output tokens} & \textbf{Input tokens (uncached)} & \textbf{Cache-read tokens} & \textbf{Cache-write tokens} \\
  \midrule
  \runid{T1_s1} & 35.9 & 4,785 & 26 & 121,305 & 24,767 \\
  \runid{T1_s2} & 48.0 & 5,582 & 28 & 169,135 & 24,056 \\
  \runid{T2_s1} & 146.0 & 15,284 & 72 & 589,775 & 55,517 \\
  \runid{T2_s2} & 65.8 & 6,849 & 34 & 241,245 & 26,715 \\
  \runid{H1_s1} & 54.4 & 10,875 & 48 & 279,835 & 36,880 \\
  \runid{H1_s2} & 52.7 & 15,447 & 52 & 306,617 & 43,184 \\
  \runid{H1_s3} & 80.4 & 14,206 & 58 & 295,628 & 62,553 \\
  \runid{H2_o1} & 48.4 & 10,544 & 38 & 358,270 & 43,040 \\
  \runid{H2_s1} & 42.8 & 4,278 & 24 & 106,313 & 16,768 \\
  \runid{H2_s2} & 21.7 & 5,561 & 26 & 120,588 & 16,792 \\
  \runid{H2_s3} & 20.3 & 3,933 & 22 & 88,801 & 15,048 \\
  \runid{H3_s1} & 32.7 & 4,264 & 22 & 90,247 & 16,071 \\
  \runid{H3_s2} & 3.6 & 4,352 & 22 & 91,068 & 15,879 \\
  \runid{H3_s3} & 23.3 & 4,512 & 22 & 88,566 & 15,195 \\
  \bottomrule
\end{tabular}
\end{center}

Scripted-arm wall-clock times are in \path{results.json} (\path{scripted.*.minutes}).

\section{Cross-check of diamond states against a published phonon-stability classifier}\label{sec:S7}

We evaluated the diamond states of this work with the 6D phonon-stability classifier of Shi et al.\ (PNAS 2024; main-text ref.~\cite{shi2024}), using its released trained weights (\path{6d_norm_stability.ckpt}, linked from \href{https://github.com/kra5h/phonon-stability}{\nolinkurl{github.com/kra5h/phonon-stability}}). The classifier was trained on DFPT-labeled diamond states (VASP, PBE) and takes the Green--Lagrange strain with tensor shear, the same convention as ours; we confirmed this from the stored energies of its training data. It reproduces the labels of its released three-dimensional normal- and shear-strain datasets with 93\% accuracy. It is an independent screen, computed with a different code and model, not a replacement for DFPT.

\begin{itemize}[leftmargin=*]
  \item \textbf{H1 agent states:} all three are classified phonon-stable (probabilities 1.00, 1.00 and 0.88). For \runid{H1_s1} this agrees with our DFPT result (lowest frequency +96.9\,\cmi{} at relaxed atomic positions; Section~\ref{sec:S10}).
  \item \textbf{Uniaxial routes (main-text Fig.~\ref{fig:physics}c):} every \dir{100}, \dir{110} and \dir{111} point is classified stable (probability $\ge 0.82$), including the \dir{111} point beyond the maximum axial stress. The minimum probability comes from an additional \dir{111} calculation at 18\% axial strain that is not shown in main-text Fig.~\ref{fig:physics}c, because its relaxation left the uniaxial branch (compressive axial stress, residual transverse stress 5.4\,GPa). The point beyond the maximum is unstable under load control, not because of a soft phonon at fixed strain, so the two results are consistent.
  \item \textbf{Diamond benchmark states and the added training states of Section~\ref{sec:S2} ($|\varepsilon_{ii}| \le 10\%$, $i = 1,2,3$):} 1,414 of 1,416 states lie inside the classifier's stability boundary; the other two (probability 0.49) are in the uncertainty-targeted set.
\end{itemize}

Results: \path{eseagent/runs/phonon_classifier_check.json}.

\section{Offline replay with the larger surrogate (exploratory)}\label{sec:S8}

To see whether the larger network of Section~\ref{sec:S2} would have changed the campaigns, we refitted it on each agent run's own DFT points and repeated the run's last \tool{optimize_strain} call unchanged. At the top candidate we computed one PBE point, full HSE06 (valley-sampled) and, for the Si candidates, the finite-strain elastic stability of main-text Section~\ref{sec:stability}. No language model was involved, and each run yields at most one candidate, so this tests the surrogate, not a new campaign.

\begin{center}
\footnotesize
\begin{tabularx}{\textwidth}{@{}>{\raggedright\arraybackslash}p{2.3cm} *{4}{>{\raggedright\arraybackslash}X}@{}}
  \toprule
  \textbf{Run} & \textbf{Agent $h$} & \textbf{Replay $h$} & \textbf{Replay \Eg, full HSE06 (eV)} & \textbf{Min.\ stiffness eigenvalue (GPa)} \\
  \midrule
  \runid{H2_o1} / s1 / s2 / s3 & 0.289 / 0.351 / 0.311 / 0.339 & 0.298 / 0.284 / 0.317 / 0.264 & 0.769 / 0.780 / 0.775 / 0.793 & 68.6 / 67.4 / 66.1 / 67.5 \\
  \runid{H3_s1} / s2 / s3 & 2.451 / 2.451 / 2.557 & 2.291 / 2.298 / 2.355 & $-0.076$ / $-0.073$ / $-0.122$ & 18.7 / 22.5 / 21.4 \\
  \runid{T1_s1} / s2 & 2.558 / 2.530 & 2.423 / 2.620 & $-0.154$ / $-0.210$ & 23.8 / 49.2 \\
  \runid{T2_s1} / s2 & 9.476 / 9.551 & 9.293 / 8.438 & 0.658 / 0.853 ($\Gamma$ CBM) & 8.0 / 21.9 \\
  \runid{H1_s1} / s3 (diamond) & 30.334 / 22.296 & 23.918 / 22.158 & 1.666 / 1.536 (target $2.0 \pm 0.03$) & not computed \\
  \bottomrule
\end{tabularx}
\end{center}

$h$ in \hunit. All 11 Si candidates meet their task targets at full HSE06, are elastically stable and have the valence-band maximum at $\Gamma$; 8 of 11 reach a lower $h$ than the agent's run (by 2--22\%). The lowest H2 state (0.264\,\hunit) lies below the H2 reference of the main text (0.281\,\hunit), and the next (0.284\,\hunit) is about 1\% above it; phonon stability was not computed for them. In diamond the replay does not help: both H1 candidates miss the 2\,eV window at full HSE06 (\runid{H1_s2} gave no feasible candidate), consistent with the scissor error, not surrogate accuracy, limiting H1. Results: \path{eseagent/runs/replay_improved.json}, \path{eseagent/runs/replay_stability.json}.

\section{Convergence settings and split statistics}\label{sec:S9}

\textbf{First-principles settings.} PseudoDojo norm-conserving pseudopotentials (main-text ref.~\cite{vansetten2018}), version 0.4, standard accuracy. The plane-wave cutoffs (50\,Ry for Si, 100\,Ry for diamond) converge total energies to better than 0.2\,meV per cell. The \kgrid{8} Monkhorst--Pack $k$-mesh is fixed in reciprocal coordinates, so it deforms with the cell; against a \kgrid{10} mesh it changes $h$ by at most 0.3\% and the gap by at most 0.4\,meV on strained Si. The DFT benchmark sets of main-text Section~\ref{sec:surr-results} were computed with the \kgrid{10} mesh.

\begin{table}[!htbp]
  \centering
  \caption{Held-out gap MAE (meV) of main-text Table~\ref{tab:mae}, mean $\pm$ s.d.\ over the five random train--test splits (60 held-out states each).}
  \label{tab:S3}
  \small
  \begin{tabularx}{\textwidth}{@{}l *{5}{>{\centering\arraybackslash}X}@{}}
    \toprule
    \textbf{Benchmark set} & \textbf{Valley-resolved (used in campaigns)} & \textbf{Direct regression} & \textbf{Band on grid (simplified)} & \textbf{Valley-resolved, 50 points} & \textbf{Valley-resolved, larger network (exploratory)} \\
    \midrule
    Si, small strain ($|\varepsilon_{ii}| \le 3\%$) & 9.2 $\pm$ 0.7 & 32.5 $\pm$ 1.5 & 34.4 $\pm$ 2.0 & 10.6 $\pm$ 1.2 & 3.8 $\pm$ 0.6 \\
    Si, wide range ($|\varepsilon_{ii}| \le 10\%$) & 38.5 $\pm$ 4.8 & 102.0 $\pm$ 7.9 & 111.1 $\pm$ 9.6 & 54.5 $\pm$ 5.2 & 24.6 $\pm$ 4.5 \\
    Diamond ($|\varepsilon_{ii}| \le 10\%$) & 123.6 $\pm$ 24.3 & 353.5 $\pm$ 42.9 & 283.2 $\pm$ 30.3 & 240.6 $\pm$ 72.8 & 77.6 $\pm$ 21.3 \\
    \bottomrule
  \end{tabularx}
\end{table}

Training points beyond 256 in Table~\ref{tab:S4} add, in a fixed random order per split, 6,200 diamond states computed after the campaigns from the same Latin-hypercube box as the diamond benchmark set ($|\varepsilon_{ii}| \le 0.10$, $|\varepsilon_{ij}| \le 0.08$ ($i \ne j$), Green--Lagrange): 300 + 500 random states and 5,400 states from overnight batches. The active-learning set is excluded. Results up to 256 points are identical to those of Table~\ref{tab:S3} and main-text Fig.~\ref{fig:surrogate}c. The added states were computed with the production \kgrid{8} $k$-mesh, whereas the diamond benchmark set (training up to 256 points and all test states) used \kgrid{10}; recomputing 12 randomly chosen added states at \kgrid{10} changed their gap by at most 6.5\,meV (mean 1.8\,meV, largest for near-metallic states), far below the model errors of about 100\,meV. The 300 and 800 added states of Section~\ref{sec:S2} used the same production mesh. All three models level off by about 1,000--2,000 points; with the network size and training schedule fixed, more data does not bring the diamond error near the 0.03\,eV H1 tolerance.

\begin{table}[!htbp]
  \centering
  \caption{Extended diamond learning curve (main-text Fig.~\ref{fig:surrogate}c, which shows up to 1,000 points): held-out gap MAE (meV), mean $\pm$ s.d.\ over the same five splits and 60 test states as Table~\ref{tab:S3}.}
  \label{tab:S4}
  \small
  \begin{tabular}{@{}rccc@{}}
    \toprule
    \textbf{Training points} & \textbf{Valley-resolved (used in campaigns)} & \textbf{Direct regression} & \textbf{Band on grid (simplified)} \\
    \midrule
    50 & 240.6 $\pm$ 72.8 & 550.0 $\pm$ 101.5 & 481.0 $\pm$ 47.5 \\
    100 & 162.0 $\pm$ 28.2 & 454.6 $\pm$ 58.2 & 336.7 $\pm$ 35.8 \\
    200 & 133.4 $\pm$ 25.1 & 370.0 $\pm$ 29.0 & 285.7 $\pm$ 28.8 \\
    256 & 123.6 $\pm$ 24.3 & 353.5 $\pm$ 42.9 & 283.2 $\pm$ 30.3 \\
    500 & 112.5 $\pm$ 18.6 & 279.5 $\pm$ 56.1 & 243.6 $\pm$ 34.6 \\
    1,000 & 102.2 $\pm$ 18.1 & 210.3 $\pm$ 25.2 & 245.3 $\pm$ 27.9 \\
    2,000 & 107.2 $\pm$ 12.7 & 180.4 $\pm$ 28.9 & 245.1 $\pm$ 24.6 \\
    4,000 & 104.1 $\pm$ 14.6 & 171.7 $\pm$ 30.5 & 250.8 $\pm$ 24.1 \\
    6,456 & 103.7 $\pm$ 14.4 & 170.7 $\pm$ 30.4 & 247.8 $\pm$ 21.4 \\
    \bottomrule
  \end{tabular}
\end{table}

Source: \path{results.json} key \path{learning_curves_ext}; script \path{eseagent/scripts/learning_curves_ext.py}.

\section{Post-campaign checks (dated amendment, 2026-10-28)}\label{sec:S10}

Three checks were added after all other results were known. The Si learning curves were later extended (Table~\ref{tab:S7}) and two checks were added after external review (Tables~\ref{tab:S8} and~\ref{tab:S9}); each is listed in Table~\ref{tab:S2} with its date. They are presented below in that order.

\textbf{Scripted arm with the naive formalization.} The scripted arm of main-text Section~\ref{sec:campaigns} was rerun once per held-out task (seed 0) with the pre-registered naive formalization (raw-PBE thresholds, full 6D strain, no stability in the search). Its tools, 150-unit budget, schedule, verification checks and single re-target were unchanged, and success was the pre-registered task criterion.

\begin{table}[!htbp]
  \centering
  \caption{Scripted arm, expert vs naive formalization.}
  \label{tab:S5}
  \small
  \begin{tabularx}{\textwidth}{@{}l c >{\raggedright\arraybackslash}X@{}}
    \toprule
    \textbf{Task} & \textbf{Expert formalization (main text)} & \textbf{Naive formalization (one seed)} \\
    \midrule
    H1 & 2/3 & 0/1: full-HSE06 gap 3.007\,eV for the first candidate; 1.933\,eV after the re-target \\
    \addlinespace
    H2 & 3/3 & 1/1, after the re-target (first candidate \hval{0.007}, full-HSE06 gap 1.171\,eV): \hval{0.36}, $\EgL{HSE06} = 0.74$\,eV and elastically stable \\
    \addlinespace
    H3 & 3/3 & 0/1: full-HSE06 gap 0.501\,eV (not metallic), $\varepsilon_{33} = -4.2\%$ (outside the film subspace); no candidate after the re-target \\
    \bottomrule
  \end{tabularx}
\end{table}

\textbf{$\bm{\Delta}$-line sampling of the full-HSE06 gap.} Full HSE06 adds the HSE06\,$-$\,PBE difference at one point of each $\Delta$ line ($s = 0.75$ of $\Gamma$--X, \kgrid{8} grid) to the PBE valley minimum (main-text Section~\ref{sec:absgap}). An HSE06 calculation on a \kgrid{12} grid ($\nq = 4$, otherwise the strict settings of main-text Section~\ref{sec:absgap}) gives the $\Delta$ correction at $s = 2/3$ and $5/6$, which bracket 0.75.

\begin{table}[!htbp]
  \centering
  \caption{Full-HSE06 gap (eV) with the $\Delta$ correction taken at three points of the $\Delta$ line.}
  \label{tab:S6}
  \small
  \begin{tabular}{@{}lccc@{}}
    \toprule
    \textbf{State} & $s = 2/3$ & $s = 0.75$ \textbf{(main text)} & $s = 5/6$ \\
    \midrule
    H1 agent state & 2.004 & 1.984 & 1.972 \\
    \dir{110} uniaxial stress, 16\% axial strain & 2.045 & 2.019 & 1.996 \\
    \dir{110} uniaxial stress, 18\% axial strain & 1.795 & 1.767 & 1.742 \\
    \bottomrule
  \end{tabular}
\end{table}

The gaps change by at most 0.03\,eV, in the same direction for all three states, so the \dir{110} comparison at the agent's gap is unchanged.

\textbf{Rule-based scoring of the formalizations.} The 22 formalization answers of main-text Section~\ref{sec:fault} were also scored by fixed text rules against the same per-task key, with no model involved. Mean score 0.945; pass counts R1 21/22, R2 18/22, R3 22/22, R4 10/10, R5 22/22. Agreement with the Sonnet and Haiku judges was 94\% and 90\%. Of the five scores on which the rules disagree with both judges, three reflect phrasings the rules miss and two concern the margin and subspace questions, on which the judges also differed for other answers.

\textbf{Extended Si learning curves (dated amendment, 2026-10-30).} After the data freeze, the Si learning curves of main-text Fig.~\ref{fig:surrogate}a,b were extended to 1,000 training points as was done for diamond (Table~\ref{tab:S4}). The added states are 770 (small strain) and 760 (wide range) Latin-hypercube points drawn with new seeds from the same strain boxes, computed with the same \kgrid{10} $k$-mesh as the Si benchmark sets; one calculation failed in each set, leaving 769 and 759 states. They are appended, in a fixed random order per split, to the training pools; the 60 test states and the five splits are those of Table~\ref{tab:S3}, and results up to the benchmark sizes (247 and 259 points) are identical to those of Table~\ref{tab:S3}. The valley-resolved error changes little beyond the benchmark size, and at 1,000 points both baselines remain less accurate than the valley-resolved model trained on 50 points (Table~\ref{tab:S3}).

\begin{table}[!htbp]
  \centering
  \caption{Extended Si learning curves (main-text Fig.~\ref{fig:surrogate}a,b): held-out gap MAE (meV), mean $\pm$ s.d.\ over the same five splits and 60 test states as Table~\ref{tab:S3}.}
  \label{tab:S7}
  \small
  \begin{tabularx}{\textwidth}{@{}l l *{3}{>{\centering\arraybackslash}X}@{}}
    \toprule
    \textbf{Benchmark set} & \textbf{Training points} & \textbf{Valley-resolved (used in campaigns)} & \textbf{Direct regression} & \textbf{Band on grid (simplified)} \\
    \midrule
    Si, small strain ($|\varepsilon_{ii}| \le 3\%$) & 247 (benchmark) & 9.2 $\pm$ 0.7 & 32.5 $\pm$ 1.5 & 34.4 $\pm$ 2.0 \\
    Si, small strain ($|\varepsilon_{ii}| \le 3\%$) & 500 & 9.0 $\pm$ 0.9 & 30.1 $\pm$ 2.9 & 34.4 $\pm$ 2.7 \\
    Si, small strain ($|\varepsilon_{ii}| \le 3\%$) & 1,000 & 8.8 $\pm$ 1.0 & 29.6 $\pm$ 2.6 & 34.0 $\pm$ 2.3 \\
    Si, small strain ($|\varepsilon_{ii}| \le 3\%$) & 1,016 (all) & 8.6 $\pm$ 1.1 & 29.5 $\pm$ 2.5 & 34.0 $\pm$ 2.2 \\
    \addlinespace
    Si, wide range ($|\varepsilon_{ii}| \le 10\%$) & 259 (benchmark) & 38.5 $\pm$ 4.8 & 102.0 $\pm$ 7.9 & 111.1 $\pm$ 9.6 \\
    Si, wide range ($|\varepsilon_{ii}| \le 10\%$) & 500 & 34.6 $\pm$ 5.1 & 92.9 $\pm$ 7.9 & 97.1 $\pm$ 9.6 \\
    Si, wide range ($|\varepsilon_{ii}| \le 10\%$) & 1,000 & 34.9 $\pm$ 2.6 & 81.0 $\pm$ 4.2 & 94.1 $\pm$ 8.0 \\
    Si, wide range ($|\varepsilon_{ii}| \le 10\%$) & 1,018 (all) & 34.8 $\pm$ 3.1 & 81.1 $\pm$ 5.3 & 94.0 $\pm$ 8.4 \\
    \bottomrule
  \end{tabularx}
\end{table}

\textbf{$\bm{\Delta}$ correction at the valley minimum (2026-10-31).} In the strained H1 states the PBE minimum of the lowest $\Delta$ valley lies at X ($s = 1$), not near $s = 0.75$. HSE06 calculations with the strict settings of main-text Section~\ref{sec:absgap} (\kgrid{8} grid, no symmetry, relaxed internal coordinates) give the $\Delta$ correction at the grid point nearest each $\Delta$ valley's PBE minimum (Table~\ref{tab:S8}). For every state this gap agrees within 0.004\,eV with the gap read directly on the HSE06 grid, except scripted H3 seed 2 ($-0.109$\,eV on the grid), which is metallic either way. Two of the three accepted H1 states fall 6--8\,meV below the $2.0 \pm 0.03$\,eV window. No other verdict among the states of Table~\ref{tab:S8} changes: the rejected or failed H1 states stay outside the window, the H1 reference stays inside it (1.994\,eV) and every Si T1 and H3 state stays metallic ($-0.142$ to $-0.252$\,eV). In the Si H2 states the $\Delta$ minimum lies near $s = 0.83$, for which $s = 0.75$ is already the nearest grid point. At the agent's gap, \dir{110} needs $h \approx 56.5$\,\hunit{} with the correction at the minimum (57.4 at $s = 0.75$), so the \dir{110} comparison is unchanged. Scoring in the main text keeps $s = 0.75$ (main-text Section~\ref{sec:absgap}).

\begin{table}[!htbp]
  \centering
  \caption{Full-HSE06 gap (eV) with the $\Delta$ correction at $s = 0.75$ and at the grid point nearest the $\Delta$ minimum.}
  \label{tab:S8}
  \small
  \begin{tabularx}{\textwidth}{@{}>{\raggedright\arraybackslash}p{4.6cm} *{3}{>{\raggedright\arraybackslash}X}@{}}
    \toprule
    \textbf{State} & $s = 0.75$ \textbf{(main text)} & \textbf{At the $\bm{\Delta}$ minimum} & \textbf{Verdict at the minimum} \\
    \midrule
    H1 agent, s1 (accepted) & 1.984 & 1.962 & outside window ($-8$\,meV) \\
    H1 agent, s2 (rejected) & 1.938 & 1.909 & unchanged \\
    H1 scripted, seed 0 (accepted) & 2.004 & 1.964 & outside window ($-6$\,meV) \\
    H1 scripted, seed 1 (failed) & 2.038 & 2.038 & unchanged \\
    H1 scripted, seed 2 (accepted; H1 reference) & 1.994 & 1.994 & unchanged \\
    H1 cBO-agent & 1.938 & 1.907 & unchanged \\
    \dir{110} uniaxial stress, 16\% / 18\% axial strain & 2.019 / 1.767 & 1.975 / 1.719 & --- \\
    Si T1 agent, s1 / s2 & $-0.208$ / $-0.188$ & $-0.248$ / $-0.227$ & unchanged \\
    Si H3 agent, s1 / s3 & $-0.166$ / $-0.212$ & $-0.205$ / $-0.252$ & unchanged \\
    Si H3 scripted, seeds 0 / 1 / 2 & $-0.107$ / $-0.108$ / $-0.142$ & $-0.146$ / $-0.148$ / $-0.142$ & unchanged \\
    Si H3 cBO-expert, seeds 0 / 1 & $-0.204$ / $-0.118$ & $-0.244$ / $-0.148$ & unchanged \\
    \bottomrule
  \end{tabularx}
\end{table}

\runid{H3_s2} replayed \runid{H3_s1} (Section~\ref{sec:S6}) and is not listed separately.

\textbf{H1 phonons at relaxed internal coordinates (2026-10-31).} The DFPT calculation for the H1 agent state (main-text Section~\ref{sec:physics}) had used the ideal two-atom basis, on which a residual force acts at this strain. Repeated at the relaxed internal coordinates (residual force $1\times10^{-4}$\,Ry\,bohr$^{-1}$) with the same settings (\kgrid{4} $q$-grid interpolated to \kgrid{16}), it gives no imaginary mode and $\wmin = +96.9$\,\cmi{} (+96.8\,\cmi{} at the ideal basis), excluding the three acoustic modes at $\Gamma$.

\begin{table}[!htbp]
  \centering
  \caption{Final structures of the T2 and H1 agent states discussed in main-text Section~\ref{sec:physics}. Green--Lagrange strain components in Voigt order with tensor shear ($\varepsilon_{11}$, $\varepsilon_{22}$, $\varepsilon_{33}$, $\varepsilon_{23}$, $\varepsilon_{13}$, $\varepsilon_{12}$), applied through $\bm{F} = (\bm{I} + 2\strain)^{1/2}$ to the zero-strain PBE primitive cell (main-text Section~\ref{sec:dft}); relaxed atomic positions in fractional coordinates of the strained cell.}
  \label{tab:S9}
  \footnotesize
  \begin{tabularx}{\textwidth}{@{}l >{\raggedright\arraybackslash}X l l@{}}
    \toprule
    \textbf{State} & $\varepsilon_{11}, \varepsilon_{22}, \varepsilon_{33}, \varepsilon_{23}, \varepsilon_{13}, \varepsilon_{12}$ & \textbf{Atom 1} & \textbf{Atom 2} \\
    \midrule
    \runid{H1_s1} (diamond) & 0.03483, $-0.07545$, 0.02960, 0.00050, 0.03687, 0.00265 & (0.00139, $-0.00135$, 0.00116) & (0.24861, 0.25135, 0.24884) \\
    \runid{T2_s1} (Si) & 0.05806, 0.05106, 0.05196, 0.05107, 0.04701, 0.04775 & (0.00073, 0.00163, 0.00146) & (0.24927, 0.24837, 0.24854) \\
    \bottomrule
  \end{tabularx}
\end{table}

Sources: \path{results.json} keys \path{scripted_naive}, \path{hse_delta_check}, \path{hse_delta_min}, \path{H1_s1_dfpt_relaxed}, \path{formalization_rule_check}; scripts \path{eseagent/scripts/scripted_arm.py} (\texttt{FORMAL = naive}), \path{hse_delta_check.py}, \path{hse_x_check.py}, \path{dfpt_h1_relaxed.py}, \path{rate_formalization_rules.py}. Table~\ref{tab:S9}: \path{results.json} key \path{agent_runs} (strain) and \path{eseagent/runs/phonon/*/pw.in} (positions). Table~\ref{tab:S7}: \path{results.json} key \path{learning_curves_ext}; scripts \path{eseagent/scripts/si_extend.py}, \path{learning_curves_ext.py}.

\end{document}